\documentclass{emulateapj}
\usepackage{apjfonts}

\def\chandra{{\it Chandra\/}}

\def\genx{{\it Generation-X\/}}
\def\einstein{{\it Einstein\/}}
\def\hst{{\it {\it HST}\/}}

\def\spitzer{{\it Spitzer\/}}

\def\xeus{{\it XEUS\/}}

\def\xmm{{\it XMM-Newton\/}}

\def\xray{\hbox{X-ray}}
\def\ecdfs{\hbox{E-CDF-S}}
\def\cdfs{\hbox{CDF-S}}
\def\cdfn{\hbox{CDF-N}}
\def\etal{{et\,al.}}

\def\ltsima{$\; \buildrel < \over \sim \;$}
\def\simlt{\lower.5ex\hbox{\ltsima}}
\def\gtsima{$\; \buildrel > \over \sim \;$}
\def\simgt{\lower.5ex\hbox{\gtsima}}
\def\kms{\ifmmode{~{\rm km~s^{-1}}}\else{~km s$^{-1}$}\fi}
\def\lsim{\lower0.3em\hbox{$\,\buildrel <\over\sim\,$}}
\def\gsim{\lower0.3em\hbox{$\,\buildrel >\over\sim\,$}}
\def\lbsol{$L_{B,\odot}$}
\def\msol{$M_\odot$}
\def\h2{H$_2$}
\def\flux{ergs~cm$^{-2}$~s$^{-1}$}
\def\xlum{ergs~s$^{-1}$}

\def\arcsec{\mbox{$^{\prime\prime}$}}
\def\arcmin{\mbox{$^\prime$}}
\def\sfr{$M_{\odot}$ yr$^{-1}$}
\def\Lx{$L_{\rm X}$}

\begin{document}

\shortauthors{LEHMER ET AL.}
\shorttitle{X-ray Study of Mass-Dependent Star-Formation History of Late-Type Galaxies}

%
\title{Tracing The Mass-Dependent Star Formation History of Late-Type Galaxies using X-ray Emission: Results from the {\it Chandra} Deep Fields.}
%

\author{
B.~D.~Lehmer,\altaffilmark{1,2}
W.~N.~Brandt,\altaffilmark{1}
D.~M.~Alexander,\altaffilmark{2}
E.~F.~Bell,\altaffilmark{3}
A.~E.~Hornschemeier,\altaffilmark{4}
D.~H.~McIntosh,\altaffilmark{5}
F.~E.~Bauer,\altaffilmark{6}
R.~Gilli,\altaffilmark{7}
V.~Mainieri,\altaffilmark{8,9}
D.~P.~Schneider,\altaffilmark{1}
J.~D.~Silverman,\altaffilmark{9,10}
A.~T.~Steffen,\altaffilmark{1}
P.~Tozzi,\altaffilmark{7}
\& C.~Wolf\altaffilmark{11}
}
\altaffiltext{1}{Department of Astronomy \& Astrophysics, 525 Davey Lab,
The Pennsylvania State University, University Park, PA 16802, USA}
\altaffiltext{2}{Department of Physics, University of Durham, South Road, Durham, DH1 3LE, UK}
\altaffiltext{3}{Max-Planck-Institut f\"ur Astronomie, K\"onigstuhl 17, D-69117 Heidelberg, Germany}
\altaffiltext{4}{Laboratory for X-ray Astrophysics, NASA Goddard Space Flight Center, Code 662, Greenbelt, MD 20771, USA}
\altaffiltext{5}{Astronomy Department, University of Massachusetts, 710 N. Pleasant St., Amherst, MA 01007, USA}
\altaffiltext{6}{Columbia Astrophysics Laboratory, Columbia University, Pupin Labortories, 550 W. 120th St., Rm 1418, New York, NY 10027, USA}
\altaffiltext{7}{Istituto Nazionale di Astrofisica (INAF) - Osservatorio Astrofisico di Arcetri, Largo E. Fermi 5, 50125 Firenze, Italy}
\altaffiltext{8}{European Southern Observatory, Karl-Schwarschild-Strasse 2, D-85748 Garching, Germany}
\altaffiltext{9}{Max-Planck-Institut f\"ur extraterrestrische Physik, Giessenbachstrasse, D-85748 Garching b. M\"unchen, Germany}
\altaffiltext{10}{Institute of Astronomy, Department of Physics, Eidgen\"ossische TechnischeHochschule, ETH Zurich, CH-8093, Switzerland}
\altaffiltext{11}{Department of Physics, University of Oxford, Keble Road, Oxford OX1 3RH, UK}

%
\begin{abstract}
%

We report on the \xray\ evolution over the last $\approx$9~Gyr of cosmic
history (i.e., since $z = 1.4$) of late-type galaxy populations in the
\chandra\ Deep Field-North and Extended \chandra\ Deep Field-South
(\hbox{CDF-N} and \hbox{E-CDF-S}, respectively; jointly CDFs) survey fields.
Our late-type galaxy sample consists of 2568 galaxies, which were identified
using rest-frame optical colors and \hst\ morphologies.  We utilized \xray\
stacking analyses to investigate the \xray\ emission from these galaxies,
emphasizing the contributions from normal galaxies that are not dominated by
active galactic nuclei (AGNs).  Over this redshift range, we find significant
increases (factors of $\approx$5--10) in the \hbox{X-ray--to--optical} mean
luminosity ratio (\Lx/$L_B$) and the \hbox{X-ray--to--stellar-mass} mean ratio
(\Lx/$M_\star$) for galaxy populations selected by $L_B$ and $M_\star$,
respectively.  When analyzing galaxy samples selected via SFR, we find that the
mean \hbox{X-ray--to--SFR} ratio ($L_{\rm X}$/SFR) is consistent with being
constant over the entire redshift range for galaxies with \hbox{SFR~=~1--100}
\sfr, thus demonstrating that \xray\ emission can be used as a robust indicator
of star-formation activity out to $z \approx 1.4$.  We find that the
star-formation activity (as traced by \xray\ luminosity) per unit stellar mass
in a given redshift bin increases with decreasing stellar mass over the
redshift range \hbox{$z =$~0.2--1}, which is consistent with previous studies
of how star-formation activity depends on stellar mass.  Finally, we extend our
\xray\ analyses to Lyman break galaxies at $z\sim3$ and estimate that \Lx/$L_B$
at $z\sim3$ is similar to its value at $z = 1.4$.

%
\end{abstract}
%


%
\section{Introduction}
%

Investigations focusing on global changes in star-formation activity and
stellar-mass build-up in field galaxies have provided significant insight into
the physical evolution of galaxies and their constituent stellar populations.
It has now been well-established that the global star-formation rate density
has declined by roughly an order of magnitude since $z \approx$~1--1.5 (e.g.,
Lilly \etal\ 1996; Madau \etal\ 1996; Steidel \etal\ 1999; Hopkins \etal\ 2004;
P{\'e}rez-Gonz{\'a}lez \etal\ 2005; Schiminovich \etal\ 2005; Colbert \etal\
2006).  Recent investigations into the details of this evolution have shown that
the star-formation history of a given galaxy depends strongly on its stellar
mass (e.g., Cowie \etal\ 1996; Juneau \etal\ 2005; Bundy \etal\ 2006; Noeske
\etal\ 2007a,b; Zheng \etal\ 2007); the peak star-formation epoch for the most
massive galaxies occured earlier in cosmic history than it did for
galaxies with lower masses. 

X-ray studies of normal late-type galaxies (i.e., those that are not dominated
by luminous active galactic nuclei [AGNs]) have shown that \xray\ emission
provides a useful, relatively-unobscured measure of star-formation activity
(e.g., Bauer \etal\ 2002a; Cohen~2003; Ranalli \etal\ 2003; Colbert \etal\
2004; Grimm \etal\ 2003; Gilfanov \etal\ 2004a; Persic \etal\ 2004; Persic \&
Rephaeli 2007; however, see Barger \etal\ 2007).  In normal galaxies, \xray\
emission originates from \xray\ binaries, supernovae, supernova remnants, hot
($\approx$\hbox{0.2--1~keV}) interstellar gas, and O-stars (see, e.g., Fabbiano~1989,
2006 for reviews).  Sensitive \chandra\ and \xmm\ studies of local late-type
galaxies have shown that high-mass \xray\ binaries (HMXBs) and low-mass \xray\
binaries (LMXBs) typically dominate the total non-nuclear \xray\ power output
(e.g., Zezas \etal\ 2002; Bauer \etal\ 2003; Soria \& Wu~2003; Swartz \etal\
2003; Jenkins \etal\ 2005; Kilgard \etal\ 2005; however, see, e.g., Doane
\etal\ 2004).  Observations indicate that the integrated \xray\ emission from
HMXB and LMXB populations trace galaxy star-formation rate (SFR) and stellar
mass ($M_\star$), respectively.  For example, using \chandra\ observations of
32 local galaxies, Colbert \etal\ (2004) found that the summed
\hbox{0.3--8~keV} non-nuclear point-source emission from a given galaxy
($L_{\rm XP}$) can be approximated as \hbox{$L_{\rm XP} \approx
\alpha$~$M_{\star}$ + $\beta$~SFR}, where $\alpha$ and $\beta$ are constants.
Therefore, galaxies having relatively high star-formation rates per unit mass
(specific star-formation rates, SSFRs) generally have dominant \xray\
point-source contributions from HMXBs (e.g., late-type star-forming galaxies),
while those with relatively low SSFRs have point-source emission primarily from
LMXBs (e.g., massive early-type galaxies).  

If the \xray\ binary populations are similarly dominating the normal-galaxy
\xray\ power output over a significant fraction of cosmic time, then there
should be a rapid increase in the globally-averaged \xray\ luminosity of normal
star-forming galaxies with cosmic look-back time in response to the increasing
global star-formation rate density (e.g., Ghosh \& White 2001).  In this
scenario, HMXBs trace the immediate star-formation rate of a galaxy and LMXBs
trace its star-formation history with a lag of a few Gyr.  With the advent of
deep \chandra\ and \xmm\ surveys (see, e.g., Brandt \& Hasinger 2005 for a
review), it has become possible to study the \xray\ properties of normal
galaxies out to $z \simgt 1$ and $z \approx 0.3$, respectively (see, e.g.,
Hornschemeier \etal\ 2000, 2003; Alexander \etal\ 2002; Georgakakis \etal\
2007; Georgantopoulos \etal\ 2005; Lehmer \etal\ 2006, 2007; Kim \etal\ 2006;
Tzanavaris \etal\ 2006; Rosa-Gonzalez \etal\ 2007).  Evidence for a global
increase in the \xray\ activity with redshift for normal late-type galaxies has
since been mounting.  Studies of the \xray\ luminosity functions of
\xray--detected normal galaxies have found that $L_{\rm X}^*$ evolves as
\hbox{$(1+z)^{1.5-3}$} over the redshift range $z \approx$~0--1.4 (e.g., Norman
\etal\ 2004; Georgakakis \etal\ 2006; Ptak \etal\ 2007; Tzanavaris \& Georgantopolous 2008).
Additionally, \xray\ stacking analyses have enabled investigations of more
representative optically-selected galaxy populations over the majority of
cosmic history ($z \approx$~0.1--4; e.g., Brandt \etal\ 2001; Hornschemeier
\etal\ 2002; Nandra \etal\ 2002; Georgakakis \etal\ 2003; Reddy \& Steidel
2004; Laird \etal\ 2005, 2006; Lehmer \etal\ 2005a).  These studies have found
that the average \xray\ luminosities of normal late-type galaxies increases
with redshift out to $z \approx$~1.4--3.  For example, using a $\approx$1~Ms
exposure of a subregion within the \chandra\ Deep Field-North, Hornschemeier
\etal\ (2002, hereafter H02) tentatively observed a factor of $\approx$2--3
increase in $L_{\rm X}/L_{B}$ from $z = 0$ to 1.4 for $L_B^*$ galaxies.
Despite these promising initial constraints, the details of the \xray\
evolution of normal late-type galaxy populations, including dependences on the
physical properties of galaxies (e.g., optical luminosity, stellar mass,
environment, and star-formation rate), have remained unexplored.

In this paper, we aim to improve significantly upon constraints for the \xray\
evolution of normal late-type galaxies (e.g., H02).  We study for the first
time how the \xray\ properties of late-type field galaxies evolve as a function
of optical luminosity, stellar mass, and star-formation rate over the redshift
range of $z =$~\hbox{0--1.4}.  We construct late-type galaxy samples located in
two of the most well-studied extragalactic \xray\ survey fields, the $\approx$2
Ms {\it Chandra} Deep Field-North (\hbox{CDF-N}; Alexander \etal\ 2003) and the
Extended \chandra\ Deep Field-South (\hbox{E-CDF-S}), which is composed of the
central $\approx$1 Ms \chandra\ Deep Field-South (CDF-S; Giacconi \etal\ 2002)
and four flanking $\approx$250~ks \chandra\ observations (Lehmer \etal\ 2005b).
These \chandra\ Deep Fields (hereafter CDFs) reach \hbox{0.5--2~keV} detection
limits of $\approx$$2.5 \times 10^{-17}$~\flux\ in the most sensitive regions
and $\simlt$$3 \times 10^{-16}$~\flux\ over the majority of the CDFs; these
levels are sufficient to detect moderately-powerful \xray\ sources ($L_{\rm
0.5-2~keV} \simgt 10^{41.5}$~\xlum) at $z = 1.4$ and $z = 0.6$, respectively.
Therefore, the CDFs comprise an unprecedented data set for effectively studying
the \xray\ emission and evolution of cosmologically distant normal galaxies
with minimal contamination from powerful AGNs.

%
%
\begin{figure*}[t]
\figurenum{1}
\centerline{
\includegraphics[width=16.0cm]{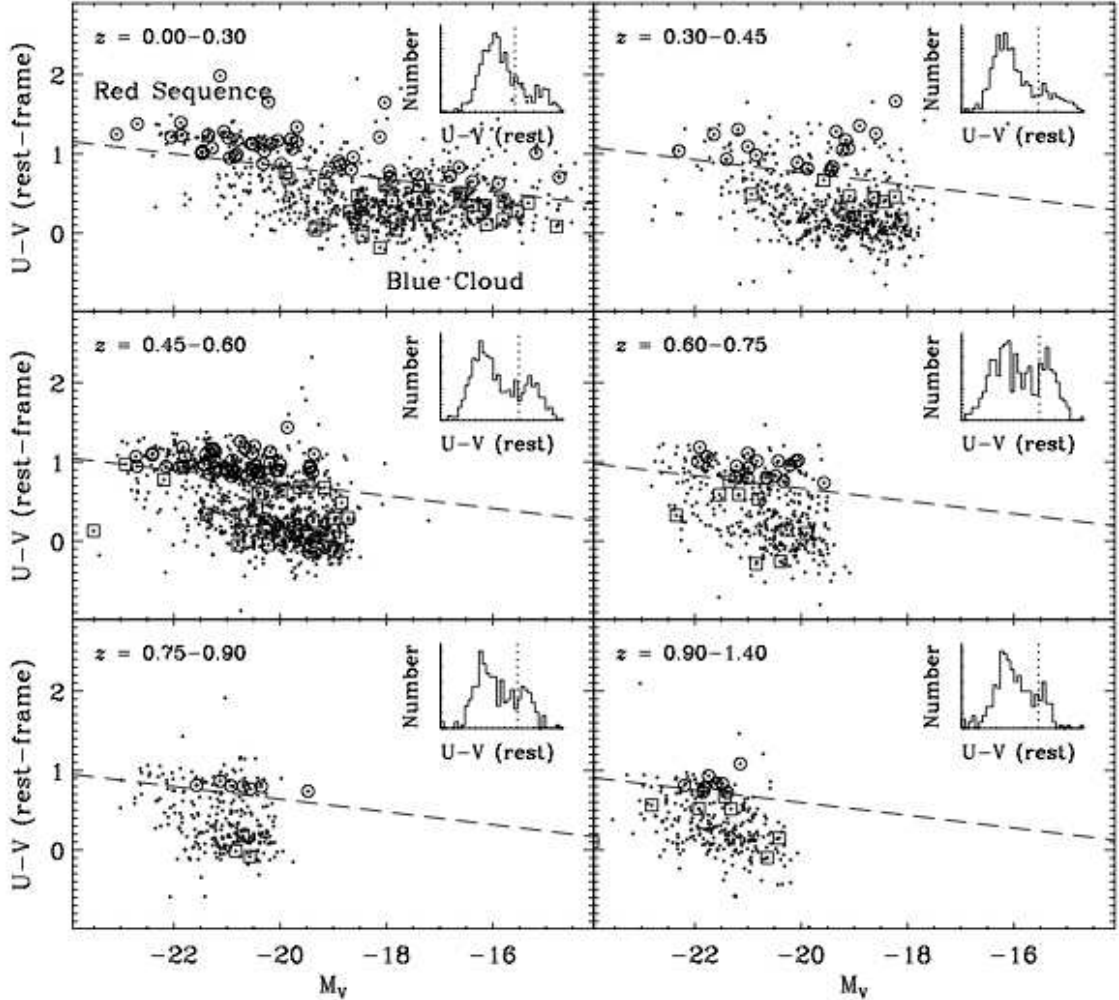}
}
\caption{
Rest-frame $U-V$ color versus absolute $V$-band magnitude $M_V$
(color-magnitude diagrams) for the 3259 $z \approx$~0--1.4 galaxies with
$z_{850} < 23$ that were within 6\farcm0 of at least one CDF aimpoint.  Each
panel shows the color-magnitude relation for a given redshift bin ({\it
annotated in the upper left-hand corners\/}).  The dashed line in each panel
represents the estimated division between red (large $U-V$ values) and blue
(small $U-V$ values) galaxy populations, which was estimated using equation~1
and the median redshift of the galaxies in each bin.  Open symbols highlight
red-sequence and blue-cloud galaxies that were visually reclassified as
late-type ({\it open circles\/}) and early-type ({\it open squares\/})
galaxies, respectively.  The inset histogram in each panel shows the
distribution of rest-frame $U-V$.  The vertical dotted line in each inset plot
indicates the estimated division between red and blue galaxy populations, which
was calculated using equation~1 and the median redshift and $M_V$ for galaxies
in each redshift bin (see $\S$~2.3).
}
\vspace{-0.15in}
\end{figure*}

The Galactic column densities are $1.3 \times 10^{20}$~cm$^{-2}$ for the \cdfn\
(Lockman 2004) and $8.8 \times 10^{19}$~cm$^{-2}$ for the \ecdfs\ (Stark \etal\
1992).  All of the \hbox{X-ray} fluxes and luminosities quoted throughout this
paper have been corrected for Galactic absorption.  Unless stated otherwise, we
quote optical magnitudes based upon the AB magnitude system for \hst\
photometry and the Vega magnitude system everywhere else.  In the \xray\ band,
we make use of three standard bandpasses: \hbox{0.5--2~keV} (soft band [SB]),
\hbox{2--8~keV} (hard band [HB]), \hbox{0.5--8~keV} (full band [FB]).
Throughout this paper, we make estimates of stellar mass and star-formation
rates using a Kroupa~(2001) initial mass function (IMF); when making
comparisons between these estimates and those quoted in other studies, we have
adjusted all values to correspond to our adopted IMF.  $H_0$ = 70~\hbox{km
s$^{-1}$ Mpc$^{-1}$}, $\Omega_{\rm M}$ = 0.3, and $\Omega_{\Lambda}$ = 0.7 are
adopted throughout this paper (e.g., Spergel \etal\ 2003), which imply a
look-back time of 7.7~Gyr at $z=1$.

%
\section{Late-Type Galaxy Sample Selection}
%

We constructed an optically selected sample of late-type galaxies within the
CDFs to use for our subsequent analyses.  We restricted our galaxy selection to
regions of the CDFs where {\it Hubble Space Telescope} (\hst) observations were
available to allow the best possible morphological classifications.  The \hst\
observations in the CDFs have been carried out via the Great Observatories
Origins Deep Survey (GOODS; Giavalisco \etal\ 2004a) and Galaxy Evolution from
Morphology and SEDs (GEMS; Rix \etal\ 2004; Caldwell \etal\ 2005) programs;
these surveys cover $\approx$90\% of the \chandra-observed regions of the CDFs
with the Advanced Camera for Surveys (ACS).  In the GOODS and GEMS regions,
photometry was available for four ($B_{435}$, $V_{606}$, $i_{775}$, and
$z_{850}$) and two ($V_{606}$ and $z_{850}$) ACS passbands, respectively.  

\subsection{Galaxy Selection Footprint}

We began building our sample by selecting all galaxies having $z_{850} < 23$.
This initial selection criterion was motivated by (1) the availability of deep
$z_{850}$ band imaging over all of the CDFs, (2) the fact that the $z_{850}$
emission probes mass-tracing rest-frame optical light redward of the 4000~\AA\ break for
galaxies at $z \simlt 1$, which constitutes the large majority of our sample,
(3) the availability of reliable spectroscopic and photometric redshifts (see
details below), and (4) the high multiwavelength detection fractions for these
sources, which allow us to determine informative rest-frame spectral energy
distributions (SEDs).  In order to isolate most effectively distant
\xray--detected AGNs, we further restricted our sample to include only galaxies
that were located in the most sensitive areas of the CDFs where the \chandra\
point-spread function (PSF) was small.  We therefore chose to include sources
having optical positions that were within 6\farcm0 of at least one of the six
\chandra\ aimpoints in the CDFs\footnote{For the CDF aimpoints, see Tables~1
and A1 of Alexander \etal\ (2003) for the $\approx$2~Ms \cdfn\ and
$\approx$1~Ms \cdfs, respectively, as well as Table~1 of Lehmer \etal\ (2005b)
for the $\approx$250~ks \ecdfs.}; the corresponding total areal footprint is
$\approx$0.18~deg$^2$.  Futhermore, we removed obvious Galactic stars that were
identified via optical spectral properties or the presence of obvious
diffraction spikes in the $z_{850}$ band images.  Under these restrictions, we
found 6905 galaxies.

\subsection{Redshifts}

We cross-correlated our initial sample of 6905 galaxies with available
spectroscopic and photometric redshift catalogs (e.g., Barger \etal\ 2003;
Le~Fevre \etal\ 2004; Szokoly \etal\ 2004; Wirth \etal\ 2004; Wolf \etal\ 2004;
Mobasher \etal\ 2004; Mignoli \etal\ 2005; Vanzella \etal\ 2005, 2006; Grazian
\etal\ 2006; Ravikumar \etal\ 2007; Silverman \etal\ 2008).  All galaxies that
did not have spectroscopic redshifts were located in the \ecdfs\ where
highly accurate (median $\delta z/1+z \approx 0.02$ for galaxies with $z_{850}
< 23$) photometric redshifts were available via COMBO-17 (Classifying Objects
by Medium-Band Observations in 17 Filters; Wolf \etal\ 2004).  In total, 6683
($\approx$97\%) of our sources had either spectroscopic or photometric
redshifts.  Visual inspection of the 222 galaxies in the \ecdfs\ without
redshifts indicated that these sources were mainly faint galaxies near bright
stars, as well as a handful of sources that were subgalactic features within
relatively nearby galaxies.  

Whenever possible, we adopted spectroscopic redshifts as the most accurate
redshifts for our galaxies.  Using the redshift information, we filtered our
sample to include only sources with $z < 1.4$ in the $\approx$2~Ms \cdfn, $z <
1$ in the $\approx$1~Ms \cdfs, and $z < 0.6$ in the $\approx$250~ks regions of
the \ecdfs; these redshift limits represent the largest distances at which we
would expect to identify moderately luminous ($L_{\rm 0.5-2~keV} \simgt
10^{41.5}$~\xlum) AGNs effectively in each respective field.   In total, 3259
galaxies remained after filtering our sample based on redshift properties.  We
used spectroscopic redshifts for 1351 ($\approx$41\%) galaxies and photometric
redshifts for the remaining 1908 ($\approx$59\%) galaxies.

\subsection{Rest-Frame Color and Morphological Selection}

The optical-color distribution for field galaxies has been shown to be bimodal,
separating ``red'' and ``blue'' galaxy populations (e.g., Strateva \etal\ 2001;
Hogg \etal\ 2002a; Blanton \etal\ 2003; Baldry \etal\ 2004).  Studies of the
color-magnitude relation for distant galaxy populations have shown that this
color bimodality is observed to persist out to at least \hbox{$z
\approx$~1--1.5} (see, e.g., Bell \etal\ 2004a; Faber \etal\ 2007; Labbe \etal\
2007), thus providing an excellent quantifiable means for separating late-type
and early-type galaxy populations.  We therefore filtered our galaxy sample to
include only sources that had blue rest-frame optical colors, as expected for
late-type galaxies that contain young stellar populations.  To estimate rest-frame $U-V$ colors
and absolute $U$, $B$, and $V$ band magnitudes ($M_U$, $M_B$, and $M_V$,
respectively) for each of the 3259 galaxies in our sample, we utilized the
available optical/near-IR data.  

For the \cdfn, we used the $U$, $B$, $V$, $R$, $I$, $z\arcmin$, and $HK\arcmin$
photometric catalogs from Capak \etal\ (2004), as well as \spitzer\ IRAC
imaging (3.6, 4.5, 5.8, and 8.0$\mu$m; Fazio \etal\ 2004) from the GOODS
(Dickinson \etal, in preparation).  For the \ecdfs, we used (1) the 17-bandpass
photometry available through COMBO-17, (2) $J$ and $K_s$ imaging from the ESO
Imaging Survey (Olsen \etal\ 2006), (3) $J$, $H$, and $K$ from the MUSYC
collaboration (Taylor \etal\ in preparation; see also, Moy \etal\ 2003; Gawiser
\etal\ 2006), and (4) \spitzer\ IRAC data from the GOODS and SIMPLE teams
(Dickinson \etal, in preparation; van~Dokkum \etal, in preparation).  Using
these data, we constructed a rest-frame near-UV--to--near-IR SED for each
galaxy.  For these SEDs, the rest-frame $U$, $B$, and $V$ filters are
well-sampled at all relevant redshifts, with the exception of sources at $z
\simgt 0.7$ where the wavelength range of the available data is sparse at
rest-frame $V$ band.  For sources at $z \simgt 0.7$, we linearly interpolated
our SED to cover the $V$ filter.  We convolved these SEDs with Johnson $U$,
$B$, and $V$ filter curves and computed rest-frame absolute magnitudes for each
respective filter following equation~5 of Hogg \etal\ (2002b).  For sources in
the \ecdfs, these computed absolute magnitudes are consistent with those
presented by Wolf \etal\ (2004).

In Figure~1 ({\it small filled circles\/}), we present rest-frame $U-V$ colors
versus $M_V$ for our galaxies in six redshift ranges.  For clarity, we also
show inset histograms giving the distribution of rest-frame $U-V$ colors for each
redshift interval.  We utilized the rest-frame $U-V$ color to divide roughly
populations of red and blue galaxies.  Use of the rest-frame $U-V$ color was
motivated by Bell \etal\ (2004a), who note that the $U$ and $V$ bandpass pair
straddle the 4000~\AA\ break, which is particularly sensitive to age and
metallicity variations of galactic stellar populations.  The dashed lines in
Figure~1 show the empirically-determined redshift-dependent color divisions
that separate blue and red galaxy populations; we calculated these divisions
following $\S$~5 of Bell \etal\ (2004a):
\begin{equation}
(U-V)_{\rm rest} = 1.15 - 0.31z - 0.08(M_V+20.7).
\end{equation}
Galaxies having rest-frame $U-V$ less than the values provided by equation~1
are often referred to as ``blue-cloud'' galaxies, while those with rest-frame
$U-V$ greater than the division are called ``red-sequence'' galaxies.  The
redshift-dependence of the blue-cloud/red-sequence galaxy division is thought
to be largely due to the evolution of the mean age and dust content of the
blue-cloud population, with a smaller contribution from changes in metallicity.  
In total, we found 2502 blue-cloud galaxies and 757 red-sequence galaxies.

Generally, the blue-cloud and red-sequence populations are composed of
late-type and early-type galaxies, respectively (see, e.g., Bell \etal\ 2004b,
2005; McIntosh \etal\ 2005).  In support of this point, we have created
Figure~2, which shows the fraction of \ecdfs\ galaxies in our sample having
blue-cloud ({\it dashed histogram\/}) and red-sequence ({\it dotted
histogram\/}) colors as a function of galaxy S\'{e}rsic index $n$.  We utilized
the H{\"a}u$\ss$ssler \etal\ (2007) S\'{e}rsic indices, which were computed
using the GEMS $z_{850}$ images and the {\ttfamily GALFIT} (Peng \etal\ 2002)
two-dimensional light-profile fitting program.  Light-profile studies of large
galaxy samples have found empirically that a S\'{e}rsic cutoff of $n=2.5$ can
roughly discriminate between late-type and early-type galaxies (e.g., Blanton
\etal\ 2003; Shen \etal\ 2003; Hogg \etal\ 2004).  Galaxies with $n \simlt 2.5$
are generally late-type galaxies, while the majority of galaxies with $n \simgt 2.5$
are early-types ({\it vertical line} in Fig.~2).  Figure~2 shows
that there is reasonable agreement between late-type and early-type galaxy
populations selected using rest-frame optical colors and S\'{e}rsic indices.

%
%
\begin{figure}
\figurenum{2}
\centerline{
\includegraphics[width=9.5cm]{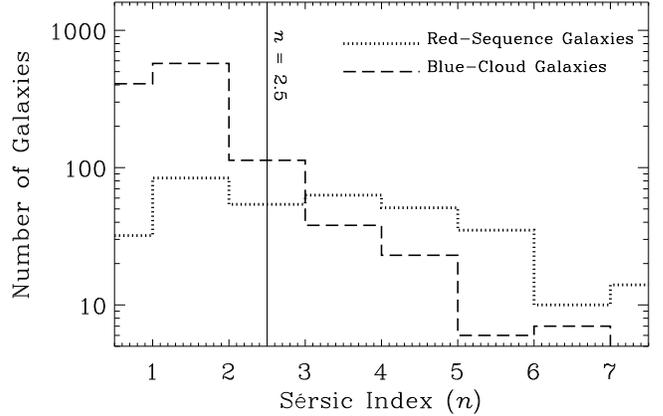}
}
\caption{
Distribution of S\'{e}rsic indices ($n$) for our initial sample of $z =$~0--1
galaxies in the \ecdfs\ having blue-cloud ({\it dashed histogram\/}) and
red-sequence ({\it dotted histogram\/}) colors.  The solid vertical line
indicates $n=2.5$, the empirical cutoff between late-type ($n \simlt 2.5$) and
early-type galaxies ($n \simgt 2.5$). 
} 
\vspace{-0.15in}
\end{figure}

%
%

\begin{figure*}
\figurenum{3}
\centerline{
\includegraphics[width=19.5cm]{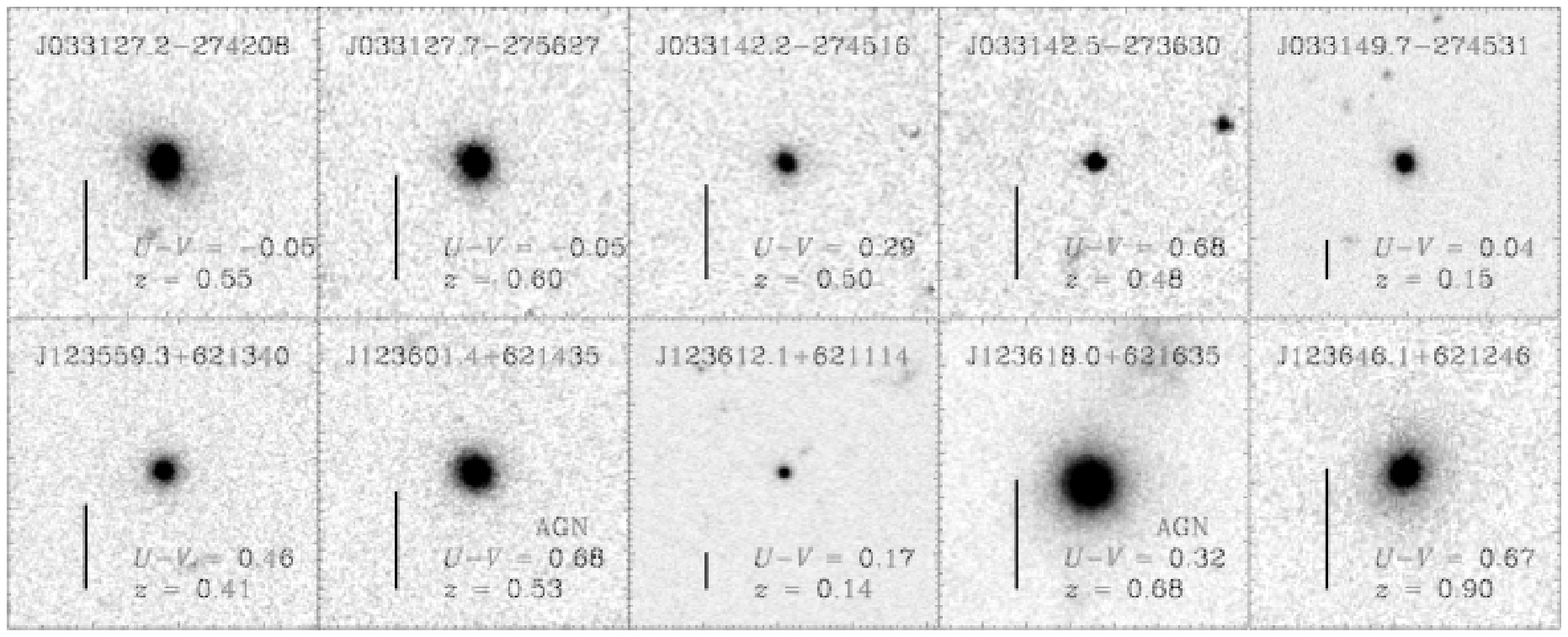}
}
\centerline{
\includegraphics[width=19.5cm]{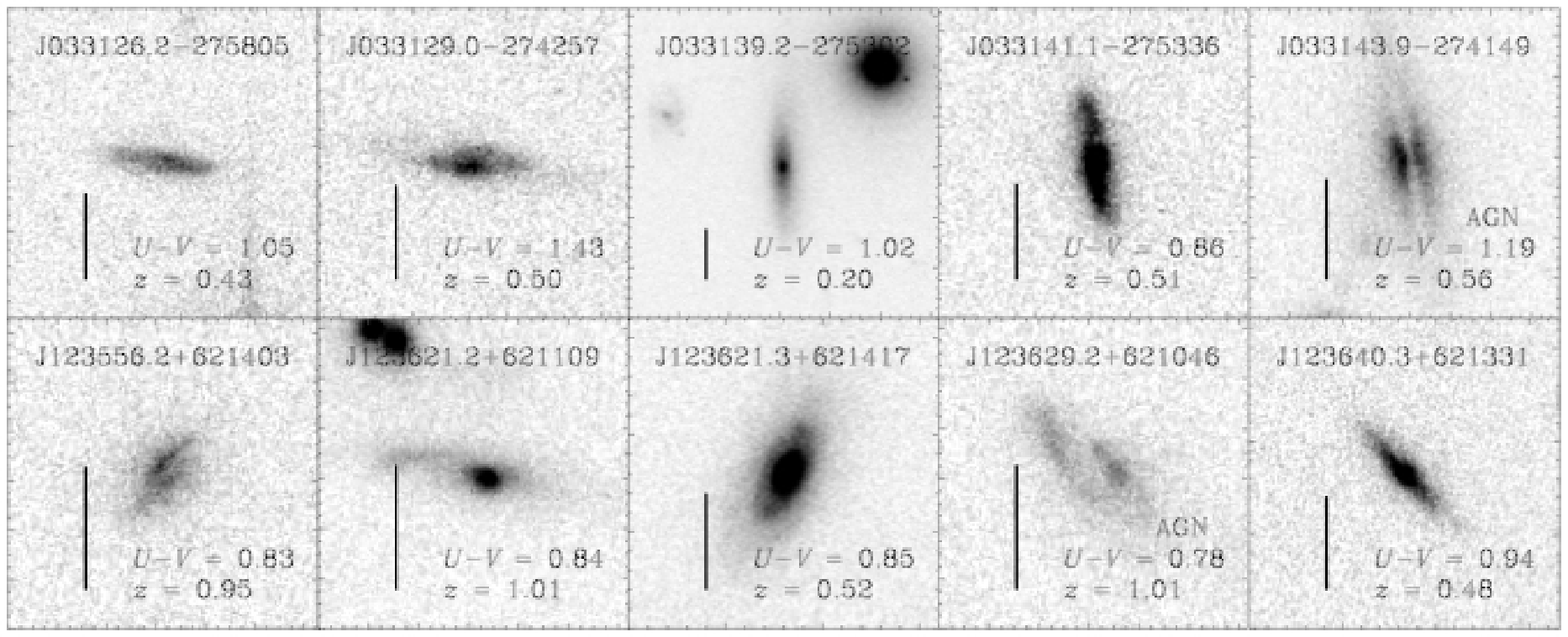}
}
\caption{
\hst\ Advanced Camera for Surveys (ACS) $z_{850}$ example images of ten
blue-cloud galaxies that have obvious early-type morphologies ({\it upper
panel\/}) and ten red-sequence galaxies that have obvious late-type
morphologies ({\it lower panel\/}); these images demonstrate the aid of the
\hst\ coverage.  Each postage-stamp image has a physical size of
30~kpc~$\times$~30~kpc, and in each image we indicate the source name ({\it
top\/}), the rest-frame $U-V$ color and redshift of the source ({\it lower
right\/}), and a vertical bar of length 1\farcs5 for scaling reference ({\it
lower left\/}); sources harboring an \xray--detected AGN (see $\S$~4.1) have
been annotated with the label ``AGN'' in the lower right. 
}
\vspace{-0.15in}
\end{figure*}

We refined further our division between late-type and early-type morphologies
by visually inspecting the $z_{850}$-band images of our entire sample of 3259
galaxies to see if there were obvious cases where the rest-frame optical colors
provided an inaccurate morphological classification.  For example, late-type
galaxies that are highly inclined to our line of sight may experience
significant reddening of the young (and blue) disk population and will have
red-sequence colors.  Also, due to variations of the stellar populations in
galaxies of a given morphological class, there will be some scatter in
rest-frame $U-V$ color near the division of the red-sequence and blue-cloud
regimes.  This will lead to a number of ``green'' galaxies that are
misclassified morphologically when the classification is based solely on
rest-frame color.  We stress that such a visual inspection is only to correct
source classifications that were {\it obviously misclassified} by simply using
color-magnitude diagrams, and our resulting sample should not be considered a
morphologically-selected sample.

Based on our visual inspection, we found 76 obvious early-type galaxies with
blue-cloud colors, and 142 obvious late-type galaxies with red-sequence colors;
we reclassified these sources as late-type and early-type galaxies,
respectively.  We inspected histograms of the fraction of blue-cloud and
red-sequence galaxies that were reclassified as a function of redshift and find
no obvious redshift-related trend.  In Figure~3, we show $z_{850}$-band
postage-stamp images of ten obvious early-type galaxies with blue-cloud colors
({\it top panels\/}), as well as ten obvious late-type galaxies with
red-sequence colors ({\it bottom panels\/}).  After reclassifying these
objects, we were left with 2568 late-type galaxies and 691 early-type galaxies
with \hbox{$z =$~0--1.4}.  Since we were interested in studying the properties
of the late-type galaxy populations, we hereafter refer to our sample of 2568
late-type galaxies as our {\it main sample}, which we use in subsequent
analyses.

\section{Physical Properties and Redshift Evolution of Late-Type Galaxies}

The primary goal of this study is to investigate the \xray\ evolution of normal
(i.e., non-AGN) late-type galaxies and to determine how this evolution depends
upon three intrinsic physical properties: optical luminosity, stellar mass, and
star-formation rate.  Below, we describe how we estimated each of these
physical properties for the galaxies in our main sample.  We note that the
populations of late-type galaxies that we are investigating here have been
selected via their intrinsic physical properties, which may have changed
significantly from the observed epoch to the present day.  For example, it is
expected that a significant fraction of the late-type galaxies in our main
sample will evolve into early-type galaxies via mergers or passive stellar
evolution (e.g., Bell \etal\ 2007).  Ideally, we would like to study the
evolution of the \xray\ properties of late-type galaxies while controlling for
such changes in the physical nature of each galaxy.  However, since the details
of this evolution are highly complex and not well understood for a given
galaxy, such a task is beyond the scope of this paper.  We therefore
investigate the \xray\ evolution of normal late-type galaxy populations in
relation to their observed intrinsic physical properties.

\subsection{Optical Luminosity}  

In order to study late-type galaxy samples selected using an observable
quantity, we made use of the $B$-band luminosity ($L_B$).  In $\S$~2.3, we
computed absolute $B$-band magnitudes $M_B$ for galaxies in our main sample
using photometrically-derived SEDs.  The $B$-band emission from a given
late-type galaxy will be significantly influenced by large populations of old
($\simgt$100~Myr) stars (measured by the stellar mass) as well as the younger
and less-numerous massive stars that reside in low-obscuration star-forming
regions (measured by the star-formation rate).  In Figure~4$a$, we show $L_B$
(expressed in solar units; where \hbox{$L_{B,\odot}= 5.2 \times
10^{32}$~\xlum}) versus redshift for galaxies in our main sample.  The
redshift-dependent selection limit of $L_B$ for our sample is set by our
$z_{850} < 23$ criterion, and at $z = 0.5$ ($z = 1.4$) this limit corresponds
to roughly \hbox{$L_B \approx 3 \times 10^{9}$~\lbsol} (\hbox{$L_B \approx 4
\times 10^{10}$~\lbsol}).  For comparison, in Figure~4$a$, we have plotted the
values of $L_B$ for the Milky Way (MW) and local galaxies M101, M82, and the
luminous infrared galaxy (LIRG) NGC~3256.  For the MW, we adopted $L_B =
1.9\times 10^{10}$~\lbsol\ as the approximate total disk-plus-bulge $B$-band
luminosity (see Table~15.2 of Gilmore \etal\ 1990).  For the local galaxies, we
utilized optical photometry from the Third Reference Catalog of Bright Galaxies
(RC3; de Vaucouleurs \etal\ 1991) and adopted distances from the IRAS Revised
Galaxy Sample (Sanders \etal\ 2003), which gives distances of 6.7, 3.6, and
35.4~Mpc and implies corresponding $B$-band luminosities of $3.3 \times
10^{10}$, $3.8 \times 10^{9}$, and $4.8 \times 10^{10}$~\lbsol\ for M101, M82,
and NGC~3256, respectively.  

\subsection{Stellar Mass}  

As discussed in $\S$~1, the \xray\ emission from LMXB populations scales with
galaxy stellar mass.  It is therefore useful to select late-type galaxies via
their stellar masses as a means for estimating the LMXB contribution to their
\xray\ emission.  To estimate the stellar mass ($M_{\star}$) of each of our
galaxies, we exploited the tight correlation between rest-frame optical color
and stellar mass-to-light ratio (e.g., Bell \& de~Jong~2001; Bell \etal\ 2003;
Kauffmann \etal\ 2003a; see also, Borch \etal\ 2006).  For this calculation, we
used a combination of rest-frame $B-V$ colors and rest-frame $K$-band
luminosities.  $K$-band luminosities were computed by fitting all available
optical/near-IR photometric data to a grid of 276 synthetic spectra generated
by the {\ttfamily P\'{E}GASE} stellar population synthesis code (Fioc \&
Rocca-Volmerange~1997).  These templates assume a single formation epoch with
an exponentially decaying star-formation history (time constants $\tau =$~1, 2,
4, and 7~Gyr) and a Kroupa \etal\ (1993) IMF; the template grid spans ages of
\hbox{0--15~Gyr}.  For each galaxy in our main sample, we convolved the
best-fit template spectrum with the $K$-band filter function to approximate the
rest-frame $K$-band luminosity, $L_K$.  This method allowed us to account for
significant curvature in the rest-frame near-IR SEDs of our sources where
photometric data points did not always overlap with the rest-frame $K$-band.
We note that 2546 ($\approx$99.1\%) of our late-type galaxies have at least one
photometric data point lying at rest-frame wavelengths between 1 and 6 $\mu$m,
which significantly constrains the SED in the rest-frame $K$-band.  We adopted
the prescription outlined in Appendix~2 of Bell \etal\ (2003) to estimate
$M_{\star}$ using rest-frame $B-V$ color and $K$-band luminosity:
\begin{equation} 
\log M_{\star}/M_{\odot} = \log L_K/L_{K,\odot} + 0.135 (B-V)
- 0.306.  
\end{equation}
\noindent The numerical constants in equation~2 were supplied by Table~7 of
Bell \etal\ (2003) and are appropriate for our choice of $B-V$ color and $L_K$;
the normalization has been adjusted by $-0.1$~dex to account for our adopted
Kroupa~(2001) IMF (see $\S$~1).  We compared our stellar-mass estimates with
those computed by Borch \etal\ (2006) for 1758 ($\approx$68\% of our main
sample) galaxies in the COMBO-17 survey and find excellent agreement between
methods, with an overall scatter of $\approx$0.2~dex.

%
%
\begin{figure}
\figurenum{4}
\centerline{
\includegraphics[width=9.cm]{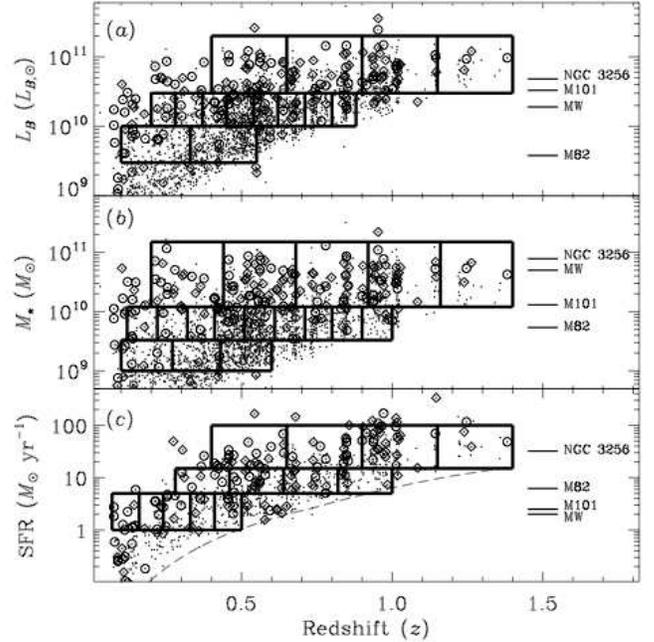}
}
\caption{
Rest-frame $B$-band luminosity $L_B$ ({\it a\/}), stellar mass $M_\star$ ({\it
b\/}), and star-formation rate SFR ({\it c\/}) versus redshift for sources in
our main sample.  For comparison, in each panel we have indicated the
properties of the MW and local galaxies M101, M82, and NGC~3256.  In
Figure~4$c$, we have included only SFRs for the 888 late-type galaxies in our
main sample that have 24$\mu$m counterparts (see $\S$~3.3 for details); the
dashed curve indicates roughly the SFR detection limit for sources without
24$\mu$m counterparts.  \xray--detected normal galaxies and AGNs have been
indicated with open circles and diamonds, respectively.  The thick gray
rectangles indicate regions where galaxy populations were selected for \xray\
stacking (see $\S\S$4.2 and 5.1).  For each galaxy. $M_\star$ and SFR were
computed following equations~2 and 3, respectively.  
} 
\vspace{-0.15in}
\end{figure}

In Figure~4$b$, we show $M_\star$ versus redshift for our sample of late-type
galaxies.  We note that our stellar-mass estimates are broadly limited by the
$z_{850} < 23$ criterion used in our sample selection.  At $z = 0.5$ ($z =
1.4$) our samples are representative for late-type galaxies with $M_{\star}
\simgt 10^{9} M_{\odot}$ ($M_{\star} \simgt 10^{10} M_{\odot}$).  For
comparison, in Figure~4$b$ we have plotted stellar masses for the MW and local
galaxies M101, M82, and NGC~3256.  We adopted a stellar mass of $5 \times
10^{10}$~$M_{\odot}$ for the MW (e.g., Hammer \etal\ 2007).  For the local
galaxies, we estimated $M_\star$ using a similar relation to equation~2, except
using $B-V$ color and $B$-band luminosity (see Appendix~2 of Bell \etal\
2003 for details); for this computation, we utilized $B-V$ colors from RC3 and
values of $L_B$ determined in $\S$~3.1.  This computation gives stellar masses
of $1.3 \times 10^{10}$, $5.4 \times 10^{9}$, and $7.8 \times
10^{10}$~$M_{\odot}$ for M101, M82, and NGC~3256, respectively.

\subsection{Star-Formation Rates}  

Since the \xray\ emission from normal late-type galaxies is known to be
strongly correlated with SFR, it is of particular interest to understand how
changes in SFR have contributed to the \xray\ evolution of the normal late-type
galaxy population.  To calculate SFRs for the galaxies in our main sample, we
utilized estimates of both the dust-uncorrected ultraviolet luminosities
($L_{\rm UV}$) originating from young stars and the infrared luminosities
(\hbox{8--1000$\mu$m}; $L_{\rm IR}$) from the dust that obscures UV light in
star-forming regions (see, e.g., Kennicutt~1998 for a review).  The former
quantity was computed following $L_{\rm UV} = 3.3 \nu l_\nu$(2800~\AA), where
$l_{\nu}$(2800 \AA) is the rest-frame 2800~\AA\ monochromatic luminosity (see
$\S$~3.2 of Bell \etal\ 2005).  $l_\nu$(2800~\AA) was approximated using
the template SEDs that were constructed in $\S$~3.2.  The latter quantity
($L_{\rm IR}$) was computed using observed-frame 24$\mu$m flux densities (i.e.,
rest-frame 24$\mu$m/$1+z$) from observations with the MIPS (Rieke \etal\ 2004)
camera onboard \spitzer.  We note that generally the dust-obscured
star-formation activity, probed here using $L_{\rm IR}$, can be measured using
either the UV spectral slope or optical nebular recombination lines (e.g.,
Balmer emission lines and \ion{O}{2}); however, the data available for our $z
\approx$~0--1.4 late-type galaxies lack the rest-frame UV and spectroscopic
coverage needed to provide such useful measurements.

Over the CDFs, deep 24$\mu$m observations were available for the GOODS fields
($f_{\rm 24 \mu m,\; lim} \approx$~30~$\mu$Jy, 6$\sigma$).\footnote{Available
at http://ssc.spitzer.caltech.edu/legacy/goodshistory.html.}  For the remaining
area (covering the outer regions of the \ecdfs), shallower 24$\mu$m
observations ($f_{\rm 24 \mu m,\; lim} \approx$~120$~\mu$Jy) were available
through SWIRE (Lonsdale \etal\ 2003).  We matched the positions of the 2568
late-type galaxies in our main sample with those from the available 24$\mu$m
source catalogs, requiring that the \hst\ and \spitzer\ centroids be offset by
no more than 1\farcs5.  We found successful matches for 888 ($\approx$35\%) of
our late-type galaxies.  For these sources, we converted 24$\mu$m flux
densities to $L_{\rm IR}$ following the methods outlined in Papovich \&
Bell~(2002).  Briefly, we utilized the entire grid of 64 infrared SEDs provided
by Dale \etal\ (2001) to estimate the mean conversion factor $\eta(z)$, which
transforms observed-frame 24$\mu$m luminosity $\nu l_\nu$(24$\mu m/1+z$) to
$L_{\rm IR}$ as a function of redshift [i.e., $L_{\rm IR} =\eta(z) \nu l_\nu(24
\mu m/1+z)$].  The mean conversion factor spans the tight range of \hbox{$\eta
=$~7.2--12.2} over the entire redshift range \hbox{$z =$~0--1.4}.  We note that
different choices of infrared SEDs yield similar results at $z \simlt 1$, but
become significantly discrepant at $z \simgt 1$ (see, e.g., Fig.~2 of Papovich
\& Bell 2002).

Using our estimates of $L_{\rm UV}$ and $L_{\rm IR}$, we calculated
star-formation rates for galaxies in our sample using the following equation:
\begin{equation} 
{\rm SFR} (M_{\odot}\;{\rm yr}^{-1}) = 9.8 \times 10^{-11}(L_{\rm IR}+L_{\rm
UV}), 
\end{equation}
\noindent where $L_{\rm IR}$ and $L_{\rm UV}$ are expressed in units of the
solar bolometric luminosity ($L_{\odot} = 3.9 \times 10^{33}$ \xlum).
Equation~3 was adopted from $\S$~3.2 of Bell \etal\ (2005) and was derived
using {\ttfamily P\'{E}GASE} stellar-population models, which assumed a 100~Myr
old population with constant SFR and a Kroupa~(2001) IMF (see Bell~2003 for
further details).  We note that for the majority of our galaxies $L_{\rm IR} >
L_{\rm UV}$, and for sources with $z \le 0.5$ and $z > 0.5$, the median
IR-to-UV ratio $L_{\rm IR}/L_{\rm UV}$~=~2.8 and 7.1, respectively.  Our $z \le
0.5$ galaxies have median $L_{\rm IR} = 1.1 \times 10^{10} L_{\odot}$, which is
characteristic of objects that are found in the nearby $D < 10$~Mpc universe.
By contrast, our $z > 0.5$ galaxies have median $L_{\rm IR} = 9.3 \times
10^{10} L_{\odot}$, similar to the local LIRG population found at
\hbox{$D>10$~Mpc}.  In Figure~4$c$, we show the distribution of SFRs for the
888 galaxies in our main sample that had 24$\mu$m counterparts.  We note that
for sources in our main sample that were within the GOODS regions, where the
24$\mu$m observations are most sensitive, the infrared detection fraction drops
from $\approx$100\% for galaxies with $z_{850} = 20 \pm 0.2$ to $\approx$20\%
for galaxies with $z_{850} = 22.8 \pm 0.2$.  This demonstrates that our SFR
completeness is limited primarily by the 24$\mu$m sensitivity limit and that
our sample of $z_{850} < 23$ late-type galaxies is highly representative of
galaxy populations above the apparent redshift-dependent SFR limit shown in
Figure~4$c$ ({\it dashed curve\/}).  At $z = 0.5$ ($z = 1.4$), this limit
corresponds to SFR~$\approx 1$~\sfr\ (SFR~$\approx 15$~\sfr).  For comparison,
in Figure~4$c$ we have plotted the SFRs for the MW and local galaxies M101,
M82, and NGC~3256.  The SFR of the MW was taken to be $\approx$2~\sfr\ (McKee
\& Williams~1997), and the SFRs of the local galaxies were computed using
equation~3.  Values of $l_\nu$(2800~\AA) were approximated from SEDs available
through the NASA/IPAC Extragalactic Database (NED)\footnote{Available at
http://nedwww.ipac.caltech.edu/.} and $L_{\rm IR}$ was taken from the IRAS
Revised Bright Galaxy Sample (Sanders \etal\ 2003).  We find SFR = 2.5, 6.2,
and 32.2~\sfr\ for M101, M82, and NGC~3256, respectively.

As a consistency check on our UV-plus-infrared SFR estimates SFR(UV+IR), we
calculated radio-derived SFRs using 1.4~GHz observations, SFR(1.4~GHz),
following equation~7 of Schmitt \etal\ (2006), which we adjusted to be
consistent with our adopted Kroupa~(2001) IMF.  We matched sources in our main
sample to 1.4~GHz catalogs, which were derived from observations using the VLA
in the \cdfn\ ($\approx$30~$\mu$Jy; Richards \etal\ 1998) and the ATCA in the
\ecdfs\ ($\approx$60~$\mu$Jy; Afonso \etal\ 2006; Rovilos \etal\ 2007).  Using
a matching radius of 1\farcs5, we found a total of 54 radio sources coincident
with our late-type galaxies.  We found that 48 ($\approx$89\%) of the
radio-detected sources had 24$\mu$m counterparts, allowing for reasonable
comparison between derived SFRs.  For these sources we found reasonable
agreement between SFRs derived from UV-plus-infrared and radio measurements,
with a mean ratio of SFR(1.4~GHz)/SFR(UV+IR)~$= 1.5 \pm 1.0$.  A large number
of late-type galaxies (840 sources) in our main sample were detected in the
24$\mu$m observations that were not detected at 1.4~GHz.  All but three of
these sources had SFR(1.4~GHz) upper limits that were consistent with that
expected from estimates of SFR(UV+IR) (see $\S$~4.1 for further details).

\section{Analysis}

\subsection{X-ray--Detected Late-Type Galaxies and AGN Identification}

We utilized the multiwavelength observations in the CDFs to obtain a census of
the active galaxies in our main sample.  We began by matching the optical
positions of our galaxies to the \xray\ positions of point sources in the CDF
catalogs of Alexander \etal\ (2003)\footnote{See
http://www.astro.psu.edu/users/niel/hdf/hdf-chandra.html for the relevant
source catalogs and data products for the $\approx$2~Ms \cdfn\ and
$\approx$1~Ms \cdfs.} for the $\approx$2~Ms \cdfn\ and $\approx$1~Ms \cdfs\ and
Lehmer \etal\ (2005b)\footnote{See
http://www.astro.psu.edu/users/niel/ecdfs/ecdfs-chandra.html for the relevant
source catalogs and data products for the $\approx$250~ks \ecdfs.} for the
$\approx$250~ks \ecdfs.  For a successful match, we required that the optical
and \xray\ centroids be displaced by no more than 1.5 times the radius of the
\chandra\ positional error circles (\hbox{80\%--90\%} confidence), which are
provided in each respective catalog.  We note that for a small number of the
galaxies in our sample at $z \simlt 0.3$, moderately luminous off-nuclear
\xray\ sources (e.g., ultraluminous \xray\ sources [ULXs]) that are
intrinsically related to the galaxies may lie outside of our adopted matching
radius; we utilized the off-nuclear \xray\ source catalog of Lehmer \etal\
(2006) to identify such galaxies and assign \xray\ properties (see details
below).

The \chandra\ source catalogs were generated using {\ttfamily wavdetect}
(Freeman \etal\ 2002) with false-positive probability thresholds of \hbox{$1
\times 10^{-7}$} and $1 \times 10^{-6}$ for the Alexander \etal\ (2003) and
Lehmer \etal\ (2005b) point-source catalogs, respectively.  However, as
demonstrated in $\S$~3.4.2 of Alexander \etal\ (2003) and $\S$~3.3.2 of Lehmer
\etal\ (2005b), legitimate lower significance \xray\ sources, detected by
running {\ttfamily wavdetect} at a false-positive probability threshold of
\hbox{$1 \times 10^{-5}$}, can be isolated by matching with relatively bright
optical sources; therefore, when matching our late-type galaxies to
\xray-detected sources, we utilized this technique.  The sky surface density
for late-type galaxies in our main sample ranges from $\approx$23000
deg$^{-2}$ in the \cdfn\ to $\approx$8800~deg$^{-2}$ in the $\approx$250~ks
\ecdfs.  The large difference between these source densities is primarily due
to differences in applied redshift cuts (see $\S$~2.2).  Given the fact that
the positional uncertainties are generally small ($\simlt$1\farcs5) for sources
within 6\farcm0 of the \chandra\ aim points, as is the case for sources in our
main sample, the corresponding estimated number of spurious matches is small.
We estimate that when using {\ttfamily wavdetect} with a false-positive
probability threshold of \hbox{$1 \times 10^{-5}$} to search for sources in
three \chandra\ bandpasses (\hbox{0.5--2~keV}, \hbox{2--8~keV}, and
\hbox{0.5--8~keV}), we expect $\approx$1.8 spurious matches.  When including
the off-nuclear sources from Lehmer \etal\ (2006), we expect an additional
$\approx$0.5 false sources; this brings our total spurious matching estimate to
$\approx$2.3 sources for our main sample.  

Using the matching criteria above, we find that 225 late-type galaxies are
detected in at least one of the \hbox{0.5--2~keV}, \hbox{2--8~keV}, or
\hbox{0.5--8~keV} bandpasses.  Out of these 225 galaxies, 12 are known
off-nuclear \xray\ sources from Lehmer \etal\ (2006).  
Since only one off-nuclear source is detected for each host galaxy, we assume
that the off-nuclear point-source dominates the total \xray\ emission from each
host galaxy.  We therefore adopted the \xray\ properties presented in Table~1
of Lehmer \etal\ (2006) for each off-nuclear host galaxy.

%
%

\begin{figure}
\figurenum{5}
\centerline{
\includegraphics[width=9.3cm]{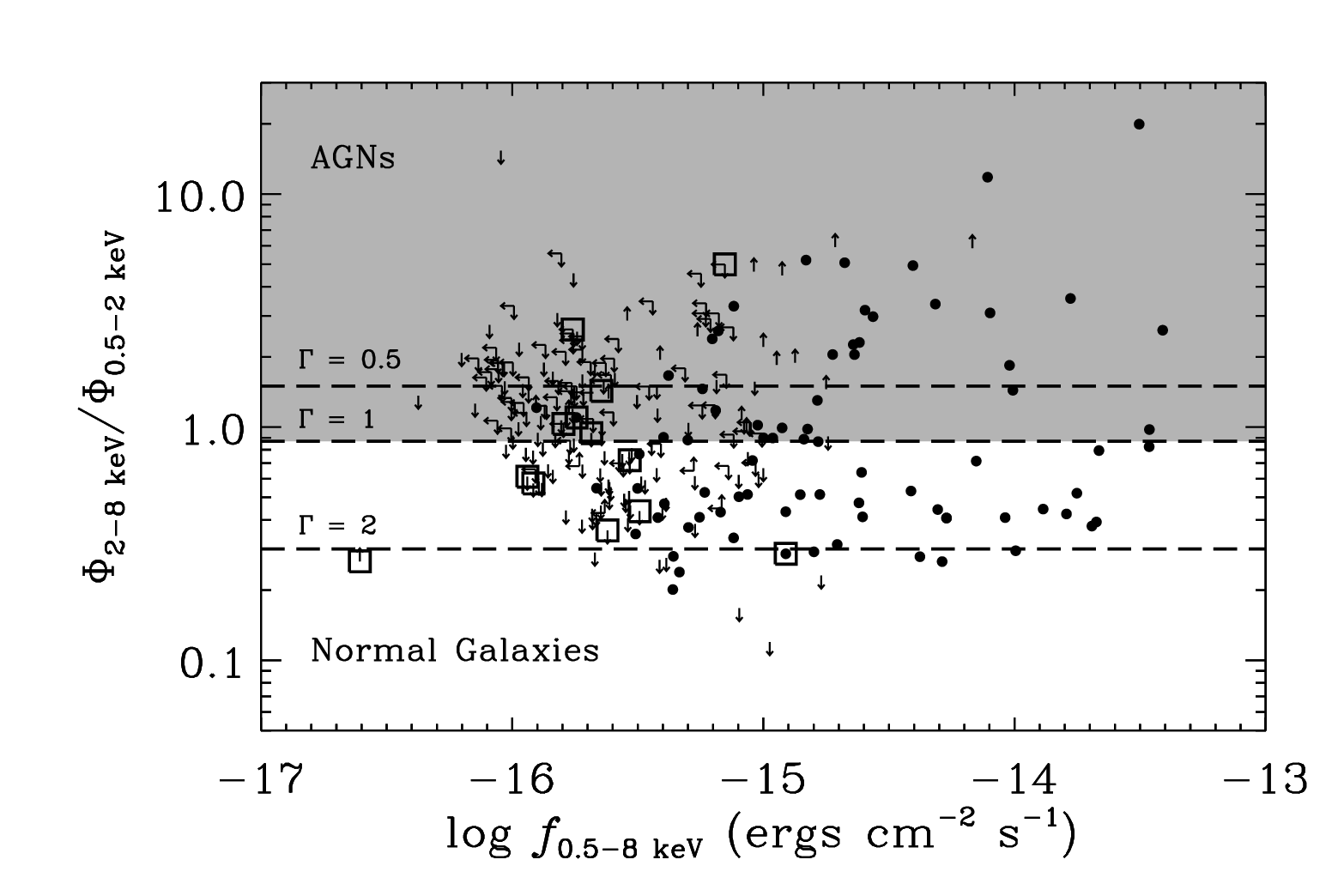}
}
\caption{
Count-rate ratio in the \hbox{2--8~keV} to \hbox{0.5--2~keV} bandpasses ($\Phi_{\rm
2-8~keV}/\Phi_{\rm 0.5-2~keV}$) versus the logarithm of the \hbox{0.5--8~keV}
flux ($\log f_{\rm 0.5-8~keV}$) for \xray--detected sources in our main sample.
Off-nuclear \xray\ sources catalogued by Lehmer \etal\ (2006) have been
highlighted with open squares.  The shaded region represents sources with
effective photon indices $\Gamma_{\rm eff} \simlt 1$; we classified these
sources as AGN candidates (see $\S$~4, criterion~1).  For reference, we have
plotted lines corresponding to $\Gamma_{\rm eff} =$~0.5, 1, and 2 ({\it dashed
lines\/}).
}
\vspace{-0.15in}
\end{figure}

%
%
\begin{figure}
\figurenum{6}
\centerline{
\includegraphics[width=9.3cm]{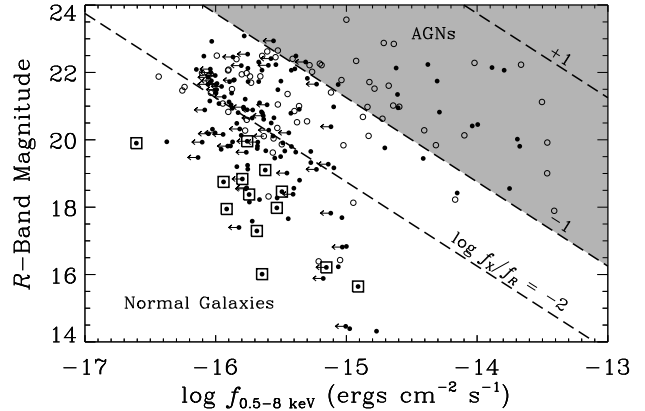}
}
\caption{
$R$-band magnitude versus $\log f_{\rm 0.5-8~keV}$ for \xray--detected sources in our
main sample.  Open circles represent sources that were classified as AGN
candidates by criterion~1 (see $\S$~4.1 and Fig.~5), and filled circles represent
all other sources; open squares have the same meaning as in Figure~5.
The shaded area represents the region where $\log(f_{\rm 0.5-8~keV}/f_{R})
> -1$; we classified sources in this region as AGN
candidates (see $\S$~4.1, criterion~2).  For reference, the dashed lines
represent $\log(f_{\rm 0.5-8~keV}/f_{R}) = -2$, $-1$, and 1.  }
\vspace{-0.15in}
\end{figure}

The unprecedented depths of the CDFs allow for the individual \xray\ detection
of luminous normal galaxies over the entire redshift range of our main sample
($z =$~0--1.4); however, the majority of the \xray--detected sources in even
the most-sensitive regions of the CDFs will be distant ($z \simgt 0.5$) AGNs
(e.g., Bauer \etal\ 2004), which we want to separate from our normal late-type
galaxy sample.  We identified AGN candidates using four primary criteria, which
utilize (1) \xray\ hardness to identify luminous obscured sources, (2)
\xray--to--optical flux ratios to identify additional relatively-unobscured
AGNs, (3) the \xray--SFR correlation to identify additional lower-luminosity
AGNs that are significantly influencing the total \xray\ emission, and (4) the
combination of radio and infrared properties to identify additional
Compton-thick AGNs and radio-loud AGNs.  As a final check on our AGN
identifications we utilized optical spectroscopic information to identify
sources with obvious AGN signatures (see criterion~5 below).  In the sections
below we provide details of each criteria.

\vspace{0.15in}

1.~{\it X-ray Hardness}.---One unique signature of moderately obscured ($N_{\rm
H} \simgt 10^{22}$~cm$^{-2}$) AGN activity is a hard \xray\ spectrum.  For
normal galaxies, the collective emission from \xray\ binaries dominates the
total \hbox{0.5--8~keV} power output.  On average, these sources have observed
power-law \xray\ SEDs with spectral slopes of \hbox{$\Gamma \approx$~1.5--1.7}
for LMXBs (e.g., Church \& Baluci{\'n}ska-Church 2001; Irwin \etal\ 2003) and
\hbox{$\Gamma \approx$~1--2} for HMXBs and ULXs (e.g., Sasaki \etal\ 2003; Liu
\& Mirabel~2005; Liu \etal\ 2006); the presence of a significant hot
interstellar gas component will steepen the resulting \xray\ spectral slope
(i.e., produce larger $\Gamma_{\rm eff}$).  To identify obscured AGNs in our
sample effectively, we flagged sources having effective photon indices of
$\Gamma_{\rm eff} \simlt 1$ as AGN candidates.  We determined $\Gamma_{\rm
eff}$ using the \hbox{2--8~keV} to \hbox{0.5--2~keV} hardness ratio $\Phi_{\rm
2-8~keV}/\Phi_{\rm 0.5-2~keV}$, where $\Phi$ is the count rate for each
bandpass.  A few sources have only \hbox{0.5--8~keV} detections.  Since these
sources were not detected in the \hbox{0.5--2~keV} bandpass, our most sensitive
bandpass, there must be a significant contribution from the \hbox{2--8~keV}
bandpass such that $\Phi_{\rm 2-8~keV}/\Phi_{\rm 0.5-2~keV} \simgt 1$
($\Gamma_{\rm eff} \simlt 0.9$).  We therefore classified these sources as AGN
candidates.  In Figure~5, we show the $\Phi_{\rm 2-8~keV}/\Phi_{\rm 0.5-2~keV}$
hardness ratio versus the logarithm of the \hbox{0.5--8~keV} flux $\log f_{\rm
0.5-8~keV}$ for the \xray--detected sources in our main sample.  The shaded
area highlights the region corresponding to $\Gamma_{\rm eff} \simlt 1$.  
Sources that had only upper limits on $\Phi_{\rm 2-8~keV}/\Phi_{\rm 0.5-2~keV}$
that lie in the shaded region were not
classified as AGN candidates, while those with lower limits on $\Phi_{\rm
2-8~keV}/\Phi_{\rm 0.5-2~keV}$ that were in
the shaded region were flagged as likely AGNs. We note that occasionally an
\xray\ luminous ULX that is too close to the nucleus of its host galaxy to be
identified as off-nuclear may have $\Gamma_{\rm eff} < 1$ and therefore be
classified as an AGN candidate via this criterion; however, since only one
identified off-nuclear \xray\ source has $\Gamma_{\rm eff} < 1$ ({\it open
squares} in Fig.~5; n.b., upper limits are not included), we do not expect such
rare sources to significantly affect our results.  Using criterion~1, we
identified a total of 71 obscured AGN candidates.

2.~{\it X-ray--to--Optical Flux Ratio}.---Detailed analyses of the \xray\
spectra of luminous AGNs in the $\approx$1~Ms \cdfs\ show that the intrinsic
AGN power-law photon index is relatively steep, \hbox{$\langle \Gamma \rangle =
1.75 \pm 0.02$} (e.g., Tozzi \etal\ 2006).  Therefore, luminous AGNs having
column densities of $N_{\rm H} \simlt 10^{22}$~cm$^{-2}$ will often have
effective photon indices of $\Gamma_{\rm eff} > 1$ and would not have been
classified as potential AGNs by criterion~1.  In order to identify luminous
AGNs with $N_{\rm H} \simlt 10^{22}$~cm$^{-2}$, we utilized the
\xray--to--optical flux ratio ($f_{\rm 0.5-8~keV}/f_{R}$) as a discriminator of
AGN activity (e.g., Maccacaro \etal\ 1988; Hornschemeier \etal\ 2000; Bauer
\etal\ 2004).  We identified sources with $\log (f_{\rm 0.5-8~keV}/f_{R}) > -1$
(see criterion~3 below for justification) as unobscured AGN candidates.  In
Figure~6, we show the $R$-band magnitude versus $\log f_{\rm 0.5-8~keV}$ for
sources in our main sample; the shaded area shows the region where $\log
(f_{\rm 0.5-8~keV}/f_{R}) > -1$.  Sources that were classified as AGN
candidates via criterion~1 are denoted with open symbols.  In total, 43
\xray--detected sources satisfied criterion~2, and 16 of these sources were
uniquely identified using this criterion (i.e., not identified by criterion~1).

%
%
\begin{figure}
\figurenum{7}
\centerline{
\includegraphics[width=9.5cm]{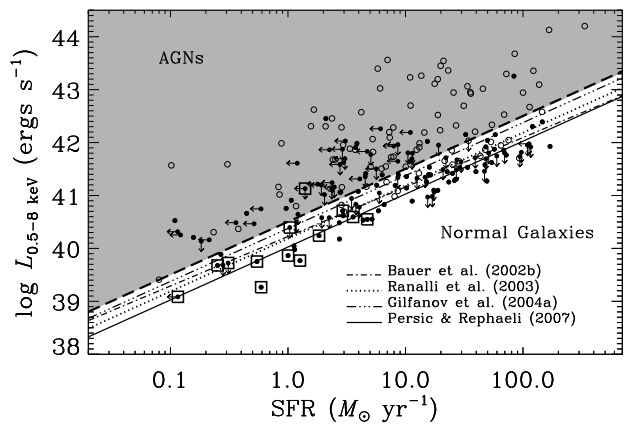}
}
\caption{
Logarithm of the \hbox{0.5--8~keV} luminosity $L_{\rm 0.5-8~keV}$ versus
SFR for \xray--detected sources in our sample that have 24$\mu$m counterparts.
Open circles represent sources that were classified as AGN candidates via
criteria~1 and 2 (see $\S$~4.1 and Figs.~5 and 6), and filled circles represent
all other sources; open squares have the same meaning as in Figure~5.
We have shown the \xray-SFR relations calibrated by Bauer \etal\ (2002b),
Ranalli \etal\ (2003), Gilfanov \etal\ (2004a), and Persic \& Rephaeli~(2007);
each respective curve has been annotated in the figure.  The shaded area above
the thick dashed line represents the region where $L_{\rm 0.5-8~keV}$ is three
times larger than its value predicted by Persic \& Rephaeli~(2007); we
classified \hbox{0.5--8~keV} detections that lie in this region as AGN
candidates (see $\S$~4.1, criterion~3).
}
\vspace{-0.15in}
\end{figure}

3.~{\it X-ray--to--SFR Correlation}.---Taken together, criteria~1 and 2 provide
an effective means for identifying AGNs that are affected by large absorption
column densities (criterion~1) and those that are notably \xray\ overluminous
for a given optical luminosity (criterion~2).  However, these criteria will
still miss moderately luminous unobscured AGNs that have $\log (f_{\rm
0.5-8~keV}/f_{R}) < -1$ (see, e.g., Peterson \etal\ 2006).  Although an
accurate classification for {\it all} such sources is currently not possible,
the situation can be mitigated using the available multiwavelength data.  We
therefore exploited the correlation between \xray\ luminosity $L_{\rm X}$ and
SFR (see $\S$~1) to identify additional AGN candidates in our main sample that
have significant \xray\ excesses over what is expected based on the \Lx-SFR
correlation.  In order to calculate the rest-frame luminosity ($L_{E_1-E_2}$;
where $E_1$ and $E_2$ are the photon-energy lower and upper bounds,
respectively) of a source having a power-law SED, we used the following
equation:
\begin{equation}
L_{E_1-E_2} = 4 \pi d^2_L f_{E_1-E_2} (1+z)^{\Gamma-2},
\end{equation} 
where $f_{E_1-E_2}$ is the observed-frame emission in the \hbox{$E_1$--$E_2$}
bandpass and $d_L$ is the luminosity distance.  Using equation~4 and an adopted
photon index of $\Gamma=2$, we calculated \hbox{0.5--8~keV} luminosities for
\xray--detected sources in our sample.  In Figure~7, we show the logarithm of
the \hbox{0.5--8~keV} luminosity $\log L_{\rm 0.5-8~keV}$ versus SFR (computed
following equation~3) for galaxies in our main sample that had 24$\mu$m
counterparts (see $\S$~3.3).  Several estimates of the \hbox{\Lx-SFR}
correlation have been shown for reference (Bauer \etal\ 2002a; Ranalli \etal\
2003; Gilfanov \etal\ 2004a; Persic \& Rephaeli~2007, hereafter PR07); these
correlations have been corrected for differences in \xray\ bandpass and SED as
well as adopted IMFs.  Hereafter, we adopt the \hbox{\Lx-SFR} correlation from
PR07 for comparisons; however, the use of other \hbox{\Lx-SFR} correlations
would yield similar results and conclusions.  Open squares show the locations
of the galaxies hosting off-nuclear sources, which appear to be preferentially
located near the \hbox{\Lx-SFR} correlation.  Open circles indicate sources
that were identified as AGN candidates via criteria~1 and 2.  Generally, these
AGNs have $L_{\rm 0.5-8~keV}$/SFR~$\simgt 3$ times that predicted by the PR07
\hbox{\Lx-SFR} correlation ({\it thick dashed line\/}), which corresponds to a
factor of $\simgt$2.5 times the RMS scatter of the PR07
\hbox{\Lx-SFR} correlation.  We therefore classified all sources in this regime
({\it shaded region\/}) as AGN candidates.  We note that sources having only
\hbox{0.5--8~keV} upper limits were not classified as AGNs.  However, sources
that were detected in the \hbox{0.5--8~keV} bandpass that had only upper limits
on the SFR were classified as AGN candidates if they were within the shaded
region of Figure~7.  Using this criterion, we identified 101 potential AGN
candidates, of which 33 were unique to criterion~3.  For the 124
\xray--detected sources that were not classified as AGNs, we found that 19
sources ($\approx$15\%) had \xray\ spectral properties indicative of AGNs
(criterion~1) and only one source ($\approx$0.8\%) had an \xray--to--optical
flux ratio elevated in comparison to normal galaxies (criterion~2).

%
%

\begin{figure}
\figurenum{8}
\centerline{
\includegraphics[width=9.5cm]{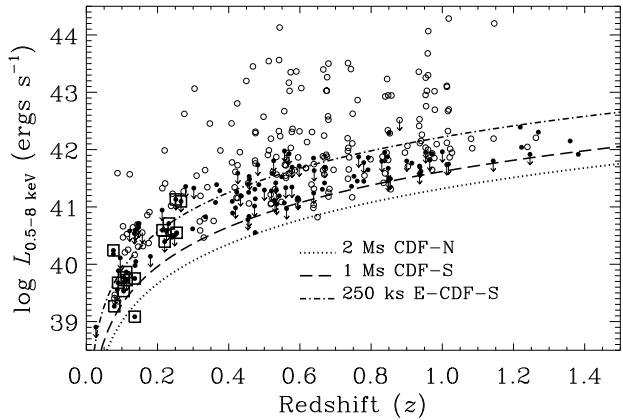}
}
\caption{
Logarithm of the \hbox{0.5--8~keV} luminosity $L_{\rm 0.5-8~keV}$ versus
redshift for the 225 \xray--detected sources in our main sample.  Open circles indicate
all of the AGN candidates that were isolated using the four criteria outlined
in $\S$~4.1, and filled circles represent potential normal galaxies; open squares
have the same meaning as in Figure~5.  The estimated \hbox{0.5--8~keV}
detection limits for the $\approx$2~Ms \cdfn, the $\approx$1~Ms \cdfs, and the
$\approx$250~ks \ecdfs\ have been indicated with dotted, dashed, and dot-dashed
curves, respectively.
}
\vspace{-0.15in}
\end{figure}

4.~{\it Infrared and Radio Properties}.---In addition to the \xray--detected
sources, we also expect there to be additional AGNs that lie below the \xray\
detection threshold that are not identified here.  In $\S$~3.3, we noted using
48 sources that the SFRs derived from the UV-plus-infrared emission are
generally consistent with those derived from radio (1.4~GHz) observations;
however, there are a few sources that are significantly scattered (i.e.,
by more than a factor of five times the intrinsic scatter) outside of the
\hbox{SFR(UV+IR)--SFR(1.4~GHz)} correlation.  These sources have either (1)
excess infrared emission due to reprocessed dust emission from a
highly-obscured ($N_{\rm H} \simgt 10^{24}$~cm$^{-2}$) moderately luminous
(intrinsic $L_{\rm X} \approx 10^{42}$--$10^{43}$~\xlum) Compton-thick AGN
(e.g., Daddi \etal\ 2007a,b) or (2) an excess of radio emission due to
prominent radio-emitting jets (e.g., Snellen \& Best~2001).  The reasonable
agreement between SFR(UV+IR) and SFR(1.4~GHz) suggests that such AGNs are not
prevalent in our late-type galaxy sample.  Ideally we would also like to
explore how UV-corrected SFRs compare to the UV-plus-infrared SFR to search for
potential Compton-thick AGNs following the approach used by Daddi \etal\
(2007a,b) for $BzK$ galaxies at $z \simgt 2$; however, since our galaxies
generally lie at much lower redshifts that the Daddi \etal\ (2007a,b) sources,
we are unable to constrain well UV spectral slopes with our available
photometry, which would be required to make dust-corrections to the observed UV
SFRs (see $\S$~3.3 for discussion).  For our sample of 24$\mu$m-detected
galaxies, we identified three radio-excess galaxies with
SFR(1.4~GHz)/SFR(UV+IR)~$>5$ and three IR-luminous galaxies with
SFR(1.4~GHz)/SFR(UV+IR)~$<1/5$.  Using these criteria we identified three new
AGN candidates, two radio luminous (\hbox{J33146.6$-$275735} and
\hbox{J33244.3$-$275141}) and one IR luminous (\hbox{J33240.0$-$274214}); one
of these sources, the radio luminous source \hbox{J33244.3$-$275141}, was
detected in the \hbox{0.5--2~keV} and \hbox{0.5--8~keV} bandpasses.   

As an additional test, we utilized IRAC photometry (see $\S$~2.3) to search for
infrared power-law sources having near-IR spectral properties characteristic of
luminous AGNs (e.g., Alonso-Herrero \etal\ 2006; Donley \etal\ 2007).  When
searching for power-law sources, we adopted the criteria discussed in Donley
\etal\ (2007).  We found that no sources in our sample satisfied these
criteria, which is consistent with the finding that most IRAC power-law sources
reside at $z \simgt 1$.   

5.~{\it Optical Spectroscopy}.---As a final check on our AGN classifications,
we searched the optical spectroscopic catalogs available for CDF sources (e.g.,
Barger \etal\ 2003; Le~Fevre \etal\ 2004; Szokoly \etal\ 2004; Wirth \etal\
2004; Vanzella \etal\ 2005, 2006) to isolate additional luminous AGNs in our
sample.  In total, we found 15 galaxies in our main sample that were classified
as AGNs via optical spectroscopy and all of these sources had been identified
as AGN candidates by the previous criteria.  We note that the majority of the
\xray--detected AGNs have moderate luminosities (intrinsic $L_{\rm X} \approx
10^{41}$--$10^{43}$~\xlum) and therefore often have high-excitation AGN
emission lines that are too faint with respect to stellar emission to be
identified via optical spectroscopy (e.g., Moran \etal\ 2002).  Therefore, it
is not surprising that we do not find any additional AGNs using this criterion.

\vspace{0.15in}

To summarize, in the \xray\ band we have detected a total of 225 ($\approx$9\%)
late-type galaxies out of the 2568 sources in our main sample.  Using the
criteria presented above, we classified 121 \xray--detected sources as AGN
candidates (with an additional two \xray--undetected AGN candidates via
criterion~4).  The remaining 104 \xray-detected sources that we do not classify
as AGN candidates are considered to be normal late-type galaxies, and we
include these galaxies in subsequent \xray\ stacking analyses (see details in
$\S$~5 below).  Thus we use 2447 late-type galaxies in our stacking analyses.
In Figure~8, we show $\log L_{\rm 0.5-8~keV}$ versus redshift for
\xray--detected sources in our main sample, and in Table~1 ({\it available
electronically\/}) we summarize their properties.  AGN candidates are denoted
with open circles and normal galaxies are plotted with filled circles.  We note
that the above criteria are not completely sufficient to classify all
\xray--detected sources that are truly AGNs as AGN candidates (see, e.g.,
Peterson \etal\ 2006). Such a misclassification is possible for low-luminosity
AGNs that are only detected in the more sensitive \hbox{0.5--2~keV} bandpass,
which have \hbox{2--8~keV} emission too weak for an accurate classification.
In $\S$~5.2, we use the \hbox{2--8~keV} AGN fraction as a function of \xray\
luminosity to argue quantitatively that we do not expect misclassified AGNs
(detected only in the \hbox{0.5--2~keV} bandpass) and low-luminosity AGNs below
the \xray\ detection limit to have a serious impact on our results.

%
%
\begin{deluxetable*}{ccccccccc}
\tabletypesize{\scriptsize}
\tablewidth{0pt}
\tablecaption{X-ray Detected Late-Type Galaxies: Source Properties}
\tablehead{
\colhead{Source Name}  &
\colhead{$z_{850}$}                     &
\colhead{}                        &
\colhead{$L_B$}                        &
\colhead{$M_\star$}                        &
\colhead{\phantom{$<$} SFR}                        &
\colhead{}                        &
\colhead{$E_{\rm 0.5-8~keV}$}                        &
\colhead{\phantom{$<$} $f_{\rm 0.5-8~keV}$}                        \\
\colhead{(J2000.0)}          &
\colhead{(mag)}                     &
\colhead{$z$}                     &
\colhead{($\log L_{B,\odot}$)}    &
\colhead{($\log M_{\odot}$)}    &
\colhead{\phantom{$<$} ($M_{\odot}$~yr$^{-1}$)}          &
\colhead{Survey}                        &
\colhead{(ks)}                        &
\colhead{\phantom{$<$} ($\log$~ergs~cm$^{-2}$~s$^{-1}$)}                        \\
\colhead{(1)}                     &
\colhead{(2)}                     &
\colhead{(3)}                     &
\colhead{(4)}                     &
\colhead{(5)}                     &
\colhead{\phantom{$<$} (6)}                     &
\colhead{(7)}                     &
\colhead{(8)}                     &
\colhead{\phantom{$<$} (9)}                     
}
\startdata
     J033132.81$-$280115.9\ldots\ldots\ldots\ldots\ldots & 17.19 &  0.15$^{\rm p}$ & 10.39 & 10.49 &              $<$5.89 & E-CDF-S 03 &  217 &              \phantom{$<$} $-$15.06 \\
     J033132.84$-$280107.5\ldots\ldots\ldots\ldots\ldots & 22.17 &  0.15$^{\rm p}$ &  8.85 &  8.44 &    \phantom{$<$}0.23 & E-CDF-S 03 &  217 &              \phantom{$<$} $-$15.45 \\
     J033139.20$-$280222.3\ldots\ldots\ldots\ldots\ldots & 17.06 &  0.25$^{\rm p}$ & 10.94 & 11.07 &    \phantom{$<$}1.40 & E-CDF-S 03 &  206 &                        $<$ $-$15.15 \\
     J033139.97$-$274157.0\ldots\ldots\ldots\ldots\ldots & 20.14 &  0.47$^{\rm s}$ & 10.14 & 10.43 &             $<$19.93 & E-CDF-S 02 &  226 &              \phantom{$<$} $-$13.46 \\
     J033141.79$-$275635.2\ldots\ldots\ldots\ldots\ldots & 19.58 &  0.51$^{\rm p}$ & 10.84 & 10.54 &             $<$13.17 & E-CDF-S 03 &  239 &              \phantom{$<$} $-$15.02 \\
\enddata
\tablecomments{Col.(1): \chandra\ source name. Col.(2): ACS $z_{850}$-band magnitude. Col.(3): Redshift estimate. Superscripts ``s'' and ``p'' indicate spectroscopic and photometric redshifts, respectively (see $\S$~2 for details). Col.(4): Logarithm of the rest-frame $B$-band luminosity in units of $L_{B,\odot}$. Col.(5): Logarithm of the stellar mass in units of $M_{\odot}$. Col.(6): Star-formation rate in units of $M_{\odot}$~yr$^{-1}$. Col.(7): Survey field in which each source was identified.  For E-CDF-S identifications, the associated field number (i.e., 01--04) indicates the \chandra\ pointing within which the source was detected (see Lehmer et~al. 2005b for details). Col.(8): Effective \hbox{0.5--8~keV} exposure time (in units of ks). Col.(9)--(11): Logarithm of the \hbox{0.5--8~keV}, \hbox{0.5--2~keV}, and \hbox{2--8~keV} flux in units of \flux. Col.(12)--(14): Logarithm of the \hbox{0.5--8~keV}, \hbox{0.5--2~keV}, and \hbox{2--8~keV} rest-frame luminosity in units of \xlum. Col.(15): Effective photon index ($\Gamma_{\rm eff}$). Col.(16): Logarithm of the \hbox{0.5--8.0~keV} to $R$-band flux ratio. Col.(17): AGN candidate (Y/N)? Col.(18): AGN selection criteria used (i.e., 1--5; see $\S$~4.1 for details).  {\it All 225 entries of Table~1 are available in the electronic edition.  A portion is shown here for guidance regarding its form and content.}}
\end{deluxetable*}
%
%

%
\begin{deluxetable*}{cccccccccc}
\tabletypesize{\scriptsize}
\tablewidth{0pt}
\tablecaption{Stacked Late-Type Normal Galaxies: Mean Properties}
\tablehead{
\colhead{}                                       &
\colhead{}                                       &
\colhead{}                                       &
\colhead{}                                       &
\multicolumn{3}{c}{Net Source Counts ($S-B$)}                   &
\multicolumn{3}{c}{Signal-to-Noise (S/N)}        \\
\colhead{}                                       &
\colhead{}                                       &
\colhead{}                                       &
\colhead{}                                       &
\multicolumn{3}{c}{\rule{1.6in}{0.01in}}                   &
\multicolumn{3}{c}{\rule{1.6in}{0.01in}}        \\
\colhead{Selection Type}                         &
\colhead{$z_{\rm mean}$}                         &
\colhead{$N_{\rm gal}$}                          &
\colhead{$N_{\rm det}$}                          &
\colhead{0.5--8~keV}                             &
\colhead{0.5--2~keV}                             &
\colhead{2--8~keV}                               &
\colhead{0.5--8~keV}                             &
\colhead{0.5--2~keV}                             &
\colhead{2--8~keV}                               \\
\colhead{(1)}                             &
\colhead{(2)}                             &
\colhead{(3)}                             &
\colhead{(4)}                             &
\colhead{(5)}                             &
\colhead{(6)}                             &
\colhead{(7)}                             &
\colhead{(8)}                             &
\colhead{(9)}                             &
\colhead{(10)}                             
}
\startdata
      $\log L_B/L_{B,\odot} =$~9.5--10.0\ldots\ldots\ldots\dots\ldots & $0.23 \pm 0.06$ & 120 &   9 &  213.3 &  158.7 &   54.9 &    8.6 &    9.6 &    3.0 \\
      $\log L_B/L_{B,\odot} =$~9.5--10.0\ldots\ldots\ldots\dots\ldots & $0.46 \pm 0.06$ & 363 &   1 &  176.7 &  152.6 &   25.3 &    4.8 &    6.8 &    0.9 \\
\\
     $\log L_B/L_{B,\odot} =$~10.0--10.5\ldots\ldots\ldots\dots\ldots & $0.24 \pm 0.02$ &  28 &   2 &   66.9 &   46.5 &   20.2 &    5.2 &    5.3 &    2.1 \\
     $\log L_B/L_{B,\odot} =$~10.0--10.5\ldots\ldots\ldots\dots\ldots & $0.33 \pm 0.03$ &  33 &   3 &   82.9 &   68.5 &   14.9 &    5.9 &    6.7 &    1.5 \\
     $\log L_B/L_{B,\odot} =$~10.0--10.5\ldots\ldots\ldots\dots\ldots & $0.41 \pm 0.03$ &  67 &   7 &  204.2 &  144.7 &   59.8 &    9.3 &    9.7 &    3.8 \\
     $\log L_B/L_{B,\odot} =$~10.0--10.5\ldots\ldots\ldots\dots\ldots & $0.50 \pm 0.02$ & 158 &   6 &  347.5 &  240.0 &  109.1 &   10.7 &   11.3 &    4.4 \\
     $\log L_B/L_{B,\odot} =$~10.0--10.5\ldots\ldots\ldots\dots\ldots & $0.57 \pm 0.02$ & 100 &   3 &  128.0 &  109.0 &   18.7 &    6.0 &    7.5 &    1.2 \\
     $\log L_B/L_{B,\odot} =$~10.0--10.5\ldots\ldots\ldots\dots\ldots & $0.67 \pm 0.02$ &  71 &   2 &  154.1 &   93.5 &   60.6 &    6.3 &    6.1 &    3.2 \\
     $\log L_B/L_{B,\odot} =$~10.0--10.5\ldots\ldots\ldots\dots\ldots & $0.75 \pm 0.02$ &  62 &   2 &  156.3 &  107.4 &   48.8 &    6.4 &    6.9 &    2.6 \\
     $\log L_B/L_{B,\odot} =$~10.0--10.5\ldots\ldots\ldots\dots\ldots & $0.84 \pm 0.02$ &  57 &   3 &  105.9 &   64.8 &   41.2 &    4.4 &    4.5 &    2.2 \\
\enddata
\tablecomments{Col.(1): Physical property (i.e., $L_B$, $M_\star$, and SFR) used to select the stacked sample. Col.(2): Mean redshift ($z_{\rm mean}$) and 1~$\sigma$ standard deviation. Col.(3): Number of galaxies stacked ($N_{\rm gal}$). Col.(4): Number of \xray--detected normal galaxies stacked. Col.(5)--(7): Net source counts ($S-B$) for the \hbox{0.5--8~keV}, \hbox{0.5--2~keV}, and \hbox{2--8~keV} bandpasses. \hbox{Col.(8)--(10)}: Signal-to-noise ratio (S/N) for the \hbox{0.5--8~keV}, \hbox{0.5--2~keV}, and \hbox{2--8~keV} bandpasses. Col.(11)--(13): Logarithm of the mean \hbox{0.5--8~keV}, \hbox{0.5--2~keV}, and \hbox{2--8~keV} flux in units of \flux. Col.(14)--(16): Fraction of the mean \hbox{0.5--8~keV}, \hbox{0.5--2~keV}, and \hbox{2--8~keV} flux originating from the \xray--undetected galaxies. Col.(17)--(19): Logarithm of the mean \hbox{0.5--8~keV}, \hbox{0.5--2~keV}, and \hbox{2--8~keV} rest-frame luminosity. Col.(20): Mean effective photon index ($\Gamma_{\rm eff}$). Col.(21): Logarithm of the \hbox{0.5--8.0~keV} to $R$-band flux ratio. Col.(22)--(24): Mean values of $L_B$, $M_\star$, and SFR for samples selected by $L_B$, $M_\star$, and SFR, respectively. Col.(25)--(27): Logarithm of the ratios $L_{\rm X}/L_B$ (ergs s$^{-1}$ $L_{B,\odot}^{-1}$), $L_{\rm X}/M_\star$ (ergs s$^{-1}$ $M_{\odot}^{-1}$), and $L_{\rm X}/$SFR [ergs~s$^{-1}$~($M_{\odot}$~yr$^{-1})^{-1}$] for samples selected by $L_B$, $M_\star$, and SFR, respectively. Col.(28)--(30): Estimated fraction of the mean \hbox{0.5--8~keV}, \hbox{0.5--2~keV}, and \hbox{2--8~keV} stacked emission originating from undetected AGNs.  {\it All 44 entries of Table~2 are available in the electronic edition.  A portion is shown here for guidance regarding its form and content.}}
\end{deluxetable*}

\subsection{X-ray Stacking Analyses of Normal Late-Type Galaxy Populations}

The majority of the normal late-type galaxies that make up our main sample
were not detected individually in the \xray\ bandpass.  In
order to study the mean \xray\ properties of these sources, we implemented
stacking analyses of galaxy populations selected by their physical properties (see $\S$~3).
We divided our main sample into subsamples (to be used for stacking) of normal
late-type galaxies selected by both physical properties (i.e., $B$-band
luminosity, stellar mass, and star-formation rate) and redshifts.  In Figure~4,
we have highlighted the divisions of our sample with thick gray rectangles,
and for normal late-type galaxies in each region, we used \xray\ stacking
analyses to constrain average properties.

For each of the subsamples defined above, we performed \xray\ stacking in each
of the three standard bandpasses (i.e., SB, HB, and FB; see $\S$~1).  We expect
these bandpasses to sample effectively power-law \xray\ emission originating
from \xray\ binaries (i.e., HMXBs and LMXBs) with a minor contribution from hot
interstellar gas in the SB for late-type galaxies at $z \simlt 0.5$.  In our
analyses, we used data products presented in Alexander \etal\ (2003) for the
$\approx$2~Ms \cdfn\ and $\approx$1~Ms \cdfs\ and Lehmer \etal\ (2005b) for the
$\approx$250~ks \ecdfs\ (see footnotes~15 and 16).  Our stacking procedure
itself was similar to that discussed in $\S$~3 of Steffen \etal\ (2007).  This
procedure differs from past stacking analyses (e.g., Lehmer \etal\ 2007) in how
the local \xray\ background of each stacked sample is determined, and produces
results that are in good agreement with the method discussed in $\S$~2.2 of
Lehmer \etal\ (2007).  For completeness, we have outlined this procedure below.

Using a circular aperture with radius $R_{\rm ap} = 1\farcs5$, we extracted
\chandra\ source-plus-background counts $S_i$ and exposure times $T_i$ (in
units of cm$^2$~s) for each galaxy using images and exposure maps,
respectively.  For a given source, we used only \chandra\ pointings with
aimpoints (see footnote~12) that were offset from the source position by less
than 6\farcm0; hereafter, we refer to this maximum offset as the inclusion
radius, $R_{\rm incl}$.\footnote{Note that $R_{\rm incl}$ has the same meaning
as it did in Lehmer \etal\ (2007).} Since the \chandra\ PSF increases in size
with off-axis angle and degrades the sensitivity for sources that are far
off-axis, our choices of $R_{\rm ap}$ and $R_{\rm incl}$ have been chosen to
give the maximal stacked signal with the majority of the PSFs being sampled by
our stacking aperture (see $\S$~2.2 and Fig.~3 of Lehmer \etal\ 2007 for
justification).  We estimate, based on stacked images of $z
\approx$~\hbox{0.1--0.3} optically luminous ($L_B > 2 \times 10^9$~\lbsol)
late-type galaxies, that $\simlt$10\% of the normal galaxy \xray\ emission
originates outside of our $1\farcs5$ radius aperture (\hbox{2.8--6.7~kpc} at
\hbox{$z =$~0.1--0.3}) for all stacked samples.  For galaxies that were within
6\farcm0 of more than one of the CDF aimpoints, we added source counts and
exposure times from all appropriate images and exposure maps, respectively;
however, there were very few sources in our main sample that met this
criterion.  

Using background maps (see $\S$~4.2 of Alexander \etal\ 2003 for the
$\approx$2~Ms \cdfn\ and $\approx$1~Ms \cdfs\ and $\S$~4 of Lehmer \etal\ 2005b
for the $\approx$250~ks \ecdfs) and exposure maps, we measured local
backgrounds $B_{i, {\rm local}}$ and exposure times $T_{i, {\rm local}}$ within
a 30~pixel~$\times$~30~pixel ($\approx$$15\arcsec \times 15\arcsec$) square, centered on
each source with the 1\farcs5 radius circle masked out.  Here again, if a
source was within 6\farcm0 of more than one of the CDF aimpoints, we summed
the local backgrounds and exposure times.  We estimated the
expected number of background counts in each circular aperture $B_i$ by scaling
the background counts within the square by the relative exposure times of the
circular aperture and the square (i.e., $B_i = B_{i, {\rm local}} \times T_i/T_{i,
{\rm local}}$).  This approach is similar to scaling the background counts in
the square by the relative areas of the circular aperture and the square; however, by
using exposure times, we are able to account more accurately for spatial
variations in pixel sensitivity due to chip gaps, bad pixels, and vignetting.
Furthermore, comparisons between this method and the Monte Carlo method used in
Lehmer \etal\ (2007) to compute $B_i$ give excellent agreement and are most
convergent for large numbers of Monte Carlo trials.

When stacking galaxy populations, we excluded sources that were (1) classified
as AGN candidates (via the criteria outlined in $\S$~4.1), (2) within 10\arcsec
of an unrelated source detected in the \xray\ catalogs, (3) within the extent
of extended \xray\ sources (see Bauer \etal\ 2002b, $\S$~3.4 of Giacconi \etal\
2002, and $\S$~6 of Lehmer \etal\ 2005b), and (4) located within 3\arcsec of
another late-type galaxy in our main sample.  We note that we include
\xray--detected normal galaxies when stacking our samples, since we are
interested in the average properties of the normal late-type galaxy population.
We tested the effects of including such sources by stacking samples both with
and without \xray--detected sources included, and find similar results for both
cases.  Typically, the \xray--detected sources contribute \hbox{$\approx$10--30\%} of
the counts in the total stacked signal.  For each of the subsamples of normal
late-type galaxies outlined in Figure~4 ({\it gray rectangles\/}), we
determined stacked source-plus-background ($S = \sum_i S_i$) and background
counts ($B = \sum_i B_i$) to determine net counts ($S-B$).  For each stacked
sample, we required that the signal-to-noise ratio [S/N~$\equiv
(S-B)/\sqrt{B}$; where $\vert S-B \vert \ll B$ and $B \ge 20$] be greater than
or equal to 3 (i.e., $\simgt$99.9\% confidence) for a detection.  For stacked
samples without significant detections, 3$\sigma$ upper limits were placed on
the source counts.

We converted the net counts obtained from each stacked sample to
absorption-corrected fluxes and rest-frame luminosities using a power-law SED
with $\Gamma = 2$.  Due to the fact that our 1\farcs5 radius stacking aperture
encircles only a fraction of the PSF\footnote{At off-axis angles $\theta
\approx 3\arcmin$, our 1\farcs5 radius circular aperture contains an
encircled-energy fraction of $\approx$100\%, 80\%, and 100\% for the SB, HB,
and FB, respectively; however at $\theta \approx 6\arcmin$, this fraction
decreases to $\approx$30\%, 25\%, and 25\%, respecively.} for sources at
relatively large off-axis angle, we calculated aperture corrections $\xi_i$ for
each stacked source $i$.  Since we are calculating average \xray\ counts from
the summed emission of many sources of differing backgrounds and exposure
times, we used a single, representative exposure-weighted aperture correction,
$\xi$.  This factor, which was determined for each stacked sample, was
calculated as follows:
\begin{equation}
\xi \equiv \frac{\sum_i \xi_i \times T_i}{T},
\end{equation}
where $T = \sum_i T_i$.  The average aperture corrections ($\xi$) for sources
in our main sample were $\approx$1.6, 1.8, and 1.7 for the SB, HB, and FB,
respectively.  Using our adopted SED, we estimated observed mean \xray\ fluxes
using the following equation:
\begin{equation}
f_{E_1-E_2} =  A_{E_1-E_2} \xi \left(\frac{S-B}{T}\right),
\end{equation}
where $A_{E_1-E_2}$ is a bandpass-dependent count-rate--to--flux conversion
factor that incorporates both the \xray\ SED information as well as Galactic
extinction using the column densities listed in $\S$~1.  These mean \xray\
fluxes were then converted to rest-frame luminosities following equation~4,
assuming a photon index of $\Gamma=2$.

%
%
\begin{figure}
\figurenum{9}
\centerline{
\includegraphics[width=9.5cm]{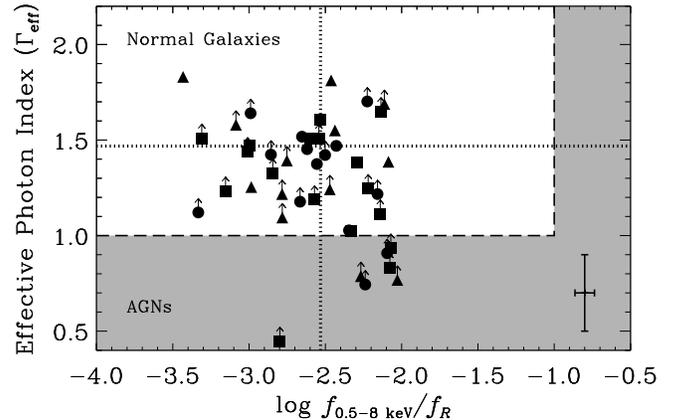}
}
\caption{
Effective photon index ($\Gamma_{\rm eff}$) versus the logarithm of the
\xray--to--optical flux ratio ($\log f_{\rm 0.5-8~keV}/f_R$) for 44 stacked
samples selected via observed properties: $L_B$ ({\it filled circles\/}),
$M_\star$ ({\it filled squares\/}), and SFR ({\it filled triangles\/}).  
The characteristic mean error bar for each quantity is given in the lower
right-hand corner. The median logarithm of the \xray--to--optical flux ratio is
indicated with a vertical dotted line ($\log f_{\rm 0.5-8~keV}/f_R = -2.53$).
The median effective photon index for the samples that were detected in both
the SB and HB is indicated with a horizontal dotted line ($\Gamma_{\rm
eff}^{\rm median} = 1.47$).  We note that all of the stacked samples with both
SB and HB detections have $\Gamma_{\rm eff} > 1$, consistent with normal
galaxies.  For the several stacked samples that have only SB detections, we
have indicated lower-limits on $\Gamma_{\rm eff}$.  The shaded regions and
corresponding boundaries ({\it dashed lines\/}) represent areas where
\xray--detected sources were classified as AGN candidates (for details, see
discussion of criteria~1 and 2 in $\S$~4.1); note that additional
criteria were used to identify potential AGNs when generating our samples of
normal late-type galaxies (see criteria~3--5 in $\S$~4.1).
}
\vspace{-0.15in}
\end{figure}

\section{Results}

\subsection{Stacking Results}

Using the stacking analysis methods discussed in $\S$~4.2, we stacked the
late-type galaxy samples presented in Figure~4 (stacked samples are denoted
with {\it thick gray rectangles\/}).  These samples were selected using $L_B$,
$M_{\star}$, and SFR, which include 14, 17, and 13 stacked samples (44 total),
respectively.  In Table~2 ({\it available electronically\/}), we tabulate our
\xray\ stacking results.  We found significant (i.e., S/N~$\simgt$~3) \xray\
detections in the \hbox{0.5--2~keV} and \hbox{0.5--8~keV} bandpasses for all
stacked samples.  In the \hbox{2--8~keV} bandpass, 15 out of the 44 stacked
samples were detected, and these samples generally constitute the most
optically luminous and massive galaxies, as well as those galaxies that are
most actively forming stars.

In Figure~9, we show the effective photon index ($\Gamma_{\rm eff}$) versus the
logarithm of the \xray--to--optical mean flux ratio ($\log f_{\rm
0.5-8~keV}/f_R$) for our 44 stacked samples ({\it filled symbols\/}) that were
selected via their observed properties.  Effective photon indices were
estimated using HB-to-SB count-rate ratios (i.e., $\Phi_{\rm 2-8~keV}/\Phi_{\rm
0.5-2~keV}$).  All 44 stacked samples have \xray--to--optical flux ratios and
\xray\ spectral slopes consistent with normal galaxies ({\it unshaded region} in
Fig.~9), suggesting that these samples are not heavily contaminated by an
underlying population of AGNs.\footnote{We note that two of our samples
have spectral slopes that are near our chosen division between normal galaxies
and AGNs ($\Gamma_{\rm eff} \sim 1$).  These samples have \hbox{$\log L_B
=$~10--10.5} and \hbox{$\log M_\star =$~9.5--10.1} and lie at $z \approx 0.66$,
a redshift where large-scale sheets of galaxies and AGNs have been isolated
previously in the \cdfs\ (e.g., Gilli \etal\ 2003)..  These galaxies have
spectral slopes that are either statistically scattered towards $\Gamma_{\rm eff}
= 1$ (see characteristic error bar in Fig.~9) or may have some
non-negligible contribution from heavily-obscured AGNs.} We find a median
logarithm of the \xray--to--optical flux ratio of $\log f_{\rm 0.5-8~keV}/f_R =
-2.53$ ({\it vertical dotted line\/}), and for samples that were detected in
both the HB and SB, the median effective photon index is $\Gamma_{\rm eff}^{\rm
median} = 1.47$ ({\it horizontal dotted line\/}).  These values are
characteristic of galaxies dominated by \xray\ binary populations.

%
%
\begin{figure}
\figurenum{10}
\centerline{
\includegraphics[width=9.1cm]{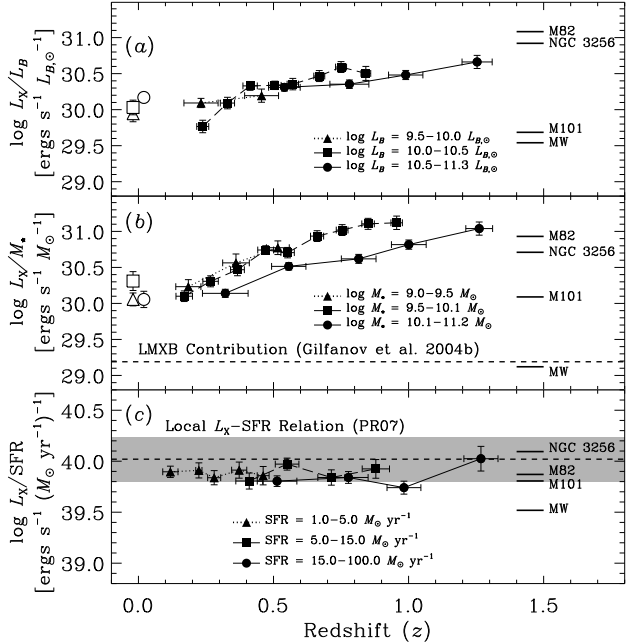}
}
\caption{
Logarithm of ({\it a\/}) the \xray--to--$B$-band mean luminosity ratio ($\log
L_{\rm X}/L_B$), ({\it b\/}) the \xray--to--stellar-mass mean ratio ($\log
L_{\rm X}/M_\star$), and ({\it c\/}) the \xray--to-star-formation-rate mean
ratio ($\log L_{\rm X}/$SFR) versus redshift ({\it filled symbols and
curves\/}) for stacked normal late-type galaxy samples selected by $L_B$
(Fig.~4$a$), $M_\star$ (Fig.~4$b$), and SFR (Fig.~4$c$), respectively.  For
comparison, in each panel we have indicated the properties of the MW and local
galaxies M101, M82, and NGC~3256.  Quoted \xray\ luminosities correspond to the
\hbox{0.5--8~keV} bandpass and were calculated following the methods described
in $\S$~4.2, assuming a power-law SED with photon index of $\Gamma = 2$.
Symbols and curves correspond to unique ranges of $L_B$, $M_\star$, and SFR,
which are annotated in each respective figure.  For reference, in Figures~10$a$
and 10$b$ we have plotted the corresponding values of $\log L_{\rm X}/L_B$ and
$\log L_{\rm X}/M_\star$, respectively, for normal late-type galaxies in the
local universe ({\it open symbols\/}) using the S01 sample.  In Figure~10$b$ we
show the expected LMXB contribution based on Gilfanov \etal\ (2004b; {\it
dashed line\/}).  Finally, in Figure~10$c$ we show the local \Lx-SFR relation
and its dispersion ({\it dashed line with shading\/}) derived by
PR07 and corrected for our choice of IMF and $\Gamma = 2$.  We note that
roughly all of our data points in Figure~10$c$ lie $\approx$\hbox{0.1--0.2} dex
below the PR07 relation, which is likely due to systematic differences
in how $L_{\rm X}$ and SFR were determined between studies.  }
\vspace{-0.15in}
\end{figure}

In Figure~10, we show the logarithm of the ratio of the \hbox{0.5--8~keV}
luminosity (hereafter, $L_{\rm X}$) to each physical property (i.e., $\log
L_{\rm X}/L_B$, $\log L_{\rm X}/M_\star$, and $\log L_{\rm X}/$SFR) versus
redshift for our samples.  Each relevant quantity for the MW and local galaxies
M101, M82, and NGC~3256 have been shown for comparison.  We adopt $6.6 \times
10^{39}$~\xlum\ as the approximate \xray\ luminosity of the MW (Grimm \etal\
2002).  For the local galaxies, we utilized the \xray\ luminosities from
Shapley \etal\ (2001; hereafter S01) for M101 ($L_{\rm X} = 1.6 \times
10^{40}$~\xlum) and M82 ($L_{\rm X} = 4.6 \times 10^{40}$~\xlum) and Lira
\etal\ (2002) for NGC~3256 ($L_{\rm X} = 4.0 \times 10^{41}$~\xlum).  We
have corrected the \xray\ luminosities to be consistent with our use of the
\hbox{0.5--8~keV} bandpass and our choice of $\Gamma = 2$.

For the purpose of comparing our results to those for nearby late-type galaxy
populations, we made use of the S01 sample of 183 normal local ($D
\simlt$~100~Mpc) spiral and irregular galaxies.  These galaxies were observed
in the \hbox{0.2--4~keV} band using the \einstein\ IPC and HRI (Fabbiano \etal\
1992) and AGNs, with luminous X-ray emission or spectral signatures indicative
of AGN activity, have been excised from the sample.  To avoid the inclusion of
early-type S0 galaxies, we chose to utilize 139 normal late-type galaxies from
the S01 sample with morphological types $T > 2$ (\hbox{Sa--Irr} Hubble types). 

The S01 sample covers ranges of $L_B$ and $M_\star$ that are well-matched to
our main sample and are representative of late-type galaxies in the local
universe.  In order to compute mean \xray\ luminosities for samples that were
directly comparable with our results, we divided the S01 sample into the same
intervals of $L_B$ and $M_\star$ that were used for our main sample (see
Figs.~4$a$ and 4$b$).  Since several of the S01 galaxies had only \xray\ upper
limits available, we computed mean \xray\ luminosities and errors using the
Kaplan-Meier estimator available through the Astronomy SURVival Analysis
software package ({\ttfamily ASURV} Rev.~1.2; Isobe \& Feigelson 1990; LaValley
\etal 1992); the Kaplan-Meier estimator handles censored data sets
appropriately.  When computing these mean \xray\ luminosities, we filtered the
S01 samples appropriately into distance intervals to avoid the Malmquist bias.
In Figures~10$a$ and 10$b$, we show the corresponding values of $\log L_{\rm
X}/L_B$ and $\log L_{\rm X}/M_\star$, respectively, for the S01 sample with
open symbols.  By contrast, the SFRs of the local sample are generally too low
($\simlt$1--10~\sfr) to provide a meaningful comparison with our distant
24$\mu$m-detected sources.  This is due to the strong positive evolution of the
star-formation rate density with redshift (see $\S$~1), which makes SFRs that
are common for galaxies in our sample ($\simgt$1--10~\sfr) comparatively rare
at $z=0$.  

From Figures~10$a$ and 10$b$, it is apparent that there is significant positive
redshift evolution in $\log L_{\rm X}/L_B$ and $\log L_{\rm X}/M_\star$ over
the redshift range of \hbox{$z \approx$~0--1.4}.  For each of the six total
selection ranges of $L_B$ and $M_\star$, the redshift progression of \xray\
luminosities is inconsistent with a constant at the $>$99.9\% confidence level.
For the most optically luminous ($L_B =$3--20~$\times 10^{10}$~\lbsol) and
massive ($M_\star =$~1--20~$\times 10^{10}$~\msol) late-type galaxies at
$z=1.4$, $L_{\rm X}/L_B$ and $L_{\rm X}/M_\star$ are measured to be larger than
the local values of S01 by factors of $3.1 \pm 0.7$ and $9.6 \pm 3.1$,
respectively.  Such values are consistent with the $\approx$$(1+z)^{1.5-3}$
evolution of $L_{\rm X}^*$ found for the normal late-type galaxy population,
which has been constrained using largely the most optically luminous and
massive galaxies (Georgakakis \etal\ 2006; Ptak \etal\ 2007; Tzanavaris \&
Georgantopoulos 2008).

The above results confirm the increase in $L_{\rm X}/L_B$ with redshift found
by H02.  In past studies, $L_{\rm X}/L_B$ has been used as a proxy for
star-formation activity (e.g., Ptak \etal\ 2001; H02; Lehmer \etal\ 2005a),
despite the fact that $L_B$ is likely to be somewhat sensitive to SFR (see
discussion in $\S$~3.1).  As discussed in $\S$~1, $L_{\rm X}$ has been shown to
be strongly correlated with galaxy SFR.  Correlation studies of spiral galaxies
in the local universe have also found strong correlations between $L_{\rm X}$
and $L_{B}$ such that $L_{\rm X} \propto L_B^{1.5}$ (S01; Fabbiano \& Shapley
2002).  The nonlinear relationship between \xray\ and $B$-band emission is
thought to be due to the increase in dust obscuration with star-formation
activity, which attenuates light from the $B$-band more effectively than it
does in the \xray\ band.  Using our sample of 47 normal late-type galaxies
with both \xray\ and 24$\mu$m detections, we found that $L_{\rm X}/L_B$ was
positively correlated with the UV dust-extinction measure ($L_{\rm IR}$+$L_{\rm
UV}$)/$L_{\rm UV}$ (Kendall's $\tau = 0.44$; $\simgt$99.99\% confidence level),
thus providing support for this hypothesis. Furthermore, it has been shown
that $L_{\rm X}/L_B$ is correlated with $L_{60 \mu {\rm m}}/L_{100 \mu {\rm
m}}$, suggesting that more intense \xray\ emission is associated with hotter IR
colors, which is indicative of intense obscured star-formation activity.
Therefore, the increase in $L_{\rm X}/L_B$ with redshift (see Fig.~10$a$)
observed for our galaxies is likely due to an increase in their star-formation
activity. 

Upon comparing $L_B$ with $M_\star$ and SFR for our galaxies, we find that
$L_B$ is well correlated with both $M_\star$ (Kendall's $\tau = 0.69$) and SFR
($\tau = 0.58$); however, the scatter in the $L_B$--$M_\star$ relation is
significantly smaller than it is for the $L_B$--SFR relation ($\approx$0.2~dex
versus $\approx$0.4~dex, respectively) implying that $L_B$ traces more
effectively $M_\star$ rather than SFR.  It is therefore not surprising that we
see similar redshift evolution of $L_{\rm X}/M_\star$ and $L_{\rm X}/L_B$. 

In Figure~10$b$, we show the estimated LMXB contribution to $\log L_{\rm
X}/M_\star$ ({\it dashed line\/}) based on Table~5 of Gilfanov \etal\ (2004b).
This value is $\approx$5--10 times lower than all mean values of $L_{\rm
X}/M_\star$, suggesting that on average LMXBs play a fairly small role in the
\xray\ emission from our stacked samples.  Furthermore, late-type galaxies in
the local universe with similar stellar masses are often found to have HMXB
emission that is $\approx$2--10 times more luminous than that expected from
LMXBs (see {\it open symbols} in Fig.~10$b$ and Fig.~3 of Gilfanov \etal\
2004c).  For galaxy samples selected via SFR, we find no evidence for
significant evolution in $\log L_{\rm X}/$SFR (Fig.~10$c$).  We note that {\it
if} the majority of the \xray\ flux from our stacked samples originated from
\xray--detected sources, then the AGN selection criteria~3 and 4 from $\S$~4.1
may potentially bias these results.  However, we find that the more populous
and unbiased \xray--undetected source populations dominate the stacked
\hbox{0.5--8~keV} signals at the $\simgt$80\% level (see column~14 of Table~2).
For each of the three ranges of SFR, $\log L_{\rm X}/$SFR is consistent with a
constant value, and has a best-fit ratio of $\log L_{\rm X}/$SFR~$ = 39.87 \pm
0.07$ ($\chi^2 = 13.1$ for 12 degrees of freedom).  These results suggest that
the contribution from LMXBs is small and that the integrated \xray\ emission
from our late-type galaxies is dominated by HMXBs.  This implies that the
evolution of our late-type galaxy samples is likely due to changes in
star-formation activity.  

%
%
%
\begin{deluxetable}{cccccc}
\tabletypesize{\scriptsize}
\tablewidth{0pt}
\tablecaption{Parametric Fitting Results For Stacked Samples}
\tablehead{
\colhead{$f_{\rm phys}$}                         &
\colhead{$\nu$}                                  &
\colhead{$\chi^2$}                               &
\colhead{$A$}                                    &
\colhead{$B$}                                    &
\colhead{$C$}                                   \\
\colhead{(1)}                             &
\colhead{(2)}                             &
\colhead{(3)}                             &
\colhead{(4)}                             &
\colhead{(5)}                             &
\colhead{(6)}                             
}
\startdata
        $L_B$\ldots\ldots & 11 &  25.37 &  $31.34 \pm 1.23$ &   $0.85 \pm 0.13$ &   $2.91 \pm 0.61$ \\
    $M_\star$\ldots\ldots & 14 &  25.73 &  $33.97 \pm 0.87$ &   $0.58 \pm 0.07$ &   $4.48 \pm 0.37$ \\
          SFR\ldots\ldots & 10 &   9.72 &  $39.91 \pm 0.06$ &   $0.89 \pm 0.11$ &   $0.32 \pm 0.88$ 
\enddata
\tablecomments{This table contains basic fitting parameters for $\chi^2$ fits to our \xray\ stacking results.  For each sample, selected by physical property $f_{\rm phys}$, we performed parametric fits for the mean \hbox{0.5--8~keV} luminosity $L_{\rm X}$ following $\log L_{\rm X} = A + B\log f_{\rm phys} + C \log (1+z)$.  Col.(1): Physical parameter $f_{\rm phys}$ used in fitting our stacking results. Col.(2): Number of degrees of freedom used in each fit. Col.(3): Minimum $\chi^2$ value for each fit. Col.(4)--(6): Best-fit values of $A$, $B$, and $C$ with errors (90\% confidence).  For futher details, see $\S$~5.1.}
\end{deluxetable}

%
%
\begin{figure}
\figurenum{11}
\centerline{
\includegraphics[width=10cm]{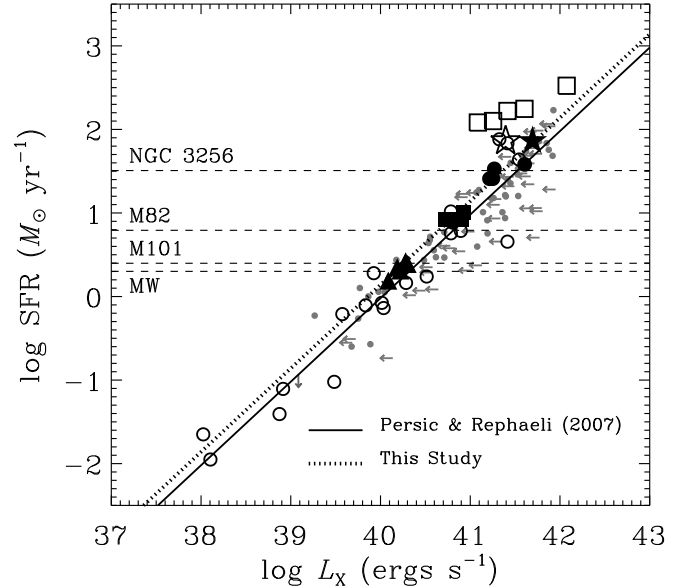}
}
\caption{
Logarithm of the star-formation rate $\log$~SFR versus the logarithm of the
\xray\ luminosity $\log L_{\rm X}$ for normal late-type galaxies.
\xray--detected sources from our main sample are indicated as small gray dots.
Sources that were not detected in the \hbox{0.5--8~keV} bandpass but were
detected in either the \hbox{0.5--2~keV} or \hbox{2--8~keV} bandpasses are
shown as upper limits.  Results from our \xray\ stacking analyses of late-type
galaxies selected via observed SFR are shown as large filled circles, squares,
and triangles, which have the same meaning as in Figure~10$c$; $z \sim 3$ LBGs
that were both uncorrected and corrected for AGN contamination have been shown
plotted as a filled star and open star, respectively.  For comparison, we show
the local galaxy sample from PR07, which includes normal late-type galaxies
({\it open circles\/}) and ultraluminous infrared galaxies (ULIRGs; {\it open
squares\/}); the best-fit PR07 relation is shown as a solid curve.  The SFRs
for the MW and local galaxies M101, M82, and NGC~3256 are indicated
({\it horizontal dashed lines\/}).  
\vspace{-0.05in}
}
\end{figure}

Since the \xray\ emission from our late-type galaxies is dominated by
star-formation processes, we note that our $M_\star$-selected stacking results
provide a relatively unobscured measure of the star-formation activity per unit
stellar mass (i.e., the SSFR; see Fig.~10$b$).  We find that at $z \approx 1$
the \xray\ emission per unit stellar mass is a factor of $\approx$\hbox{2--3}
larger for galaxies with \hbox{$M_\star =$~3--10~$\times 10^9$~\msol} versus
that observed for galaxies with \hbox{$M_\star =$~1--20~$\times
10^{10}$~\msol}.  At $z \approx$~1, we find that $L_{\rm X}/M_{\star}$ is
larger than its local value (S01) by factors of $6.4\pm2.2$ and $5.8\pm1.6$ for
late-type galaxies with \hbox{$M_\star =$~3--10~$\times 10^9$~\msol} and
\hbox{$M_\star =$~1--20~$\times 10^{10}$~\msol}, respectively.  These results
are broadly consistent with observed differences in the mean SSFRs found by
Zheng \etal\ (2007) for $z \approx 1$ galaxies of comparable stellar masses,
and imply that the lower-mass galaxies are undergoing more significant stellar
mass growth over \hbox{$z \approx$~0--1} than more massive galaxies.

In order to quantify the dependences of the \xray\ luminosity on redshift and
physical properties, we performed multivariate parametric fitting to our
stacked data.  For each galaxy sample selected via $L_B$, $M_\star$, and SFR,
we investigated the redshift evolution of the \xray\ luminosity.  For this
analysis, we fit our data to a power-law parametric form:
\begin{equation}
\log L_{\rm X}(f_{\rm phys},z) = A + B \log f_{\rm phys} + C \log (1+z),
\end{equation}
where $f_{\rm phys}$ is a place holder for each of the three physical
properties ($L_B$, $M_\star$, and SFR) used for our sample selections, and $A$,
$B$, and $C$ are fitting constants.  For each sample, we utilized our \xray\
stacking results, equation~7, and $\chi^2$ fitting to estimate the constants
$A$, $B$, and $C$.  For our adopted three-component model, we constrained $A$,
$B$, and $C$ using 90\% confidence errors ($\Delta \chi^2 = 2.7$).  The S01
local data points were not used for these fits due to differences in galaxy
selection, instrument calibration, and AGN identification.  

%
%
\begin{figure}
\figurenum{12}
\centerline{
\includegraphics[width=9.3cm]{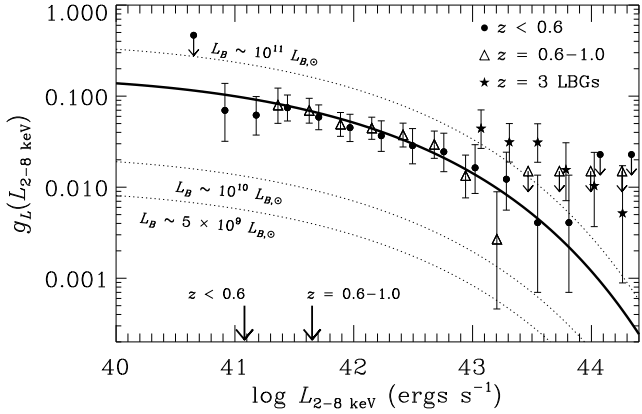}
}
\caption{
Cumulative \xray\ luminosity dependent AGN fraction $g_L(L_{\rm 2-8~keV})$
versus $L_{\rm 2-8~keV}$ for late-type galaxies with $L_B \simgt 2 \times
10^{10}$~\lbsol\ at $z \simlt 0.6$ ({\it filled circles\/}) and $z
\approx$~0.6--1 ({\it open triangles\/}).  For reference, we have indicated the
corresponding AGN fraction for Lyman break galaxies at $z \sim 3$ ({\it filled
stars\/}; see $\S$~5 for further details).  We have indicated the median \xray\
detection limit for galaxies in each redshift range ({\it downward-pointing
arrows along the x-axis\/}).  The solid curve represents the best-fit relation
for $g_L(L_{\rm 2-8~keV})$, which was fit using all late-type galaxies with $z
\approx$~0--1 and $L_B \simgt 2 \times10^{10}$~\lbsol\ in our main sample.  For
reference, we have shown the estimated AGN fraction for galaxies with $L_B
\approx 5 \times 10^9$, $10^{10}$, and $10^{11}$~\lbsol\ ({\it dotted
curves\/}; see $\S$~5.2 for further details). 
\vspace{-0.15in}
}
\end{figure}

In Table~3, we tabulate our constraints on $\chi^2$, $A$, $B$, and $C$ for
these fits.  We find that this particular choice (i.e., eqn.~7) of
parameterization does not provide acceptable fits for galaxy samples selected
via $L_B$ and $M_\star$.  However, for galaxy samples selected via SFR, we find
a good fit for this parameterization ($\chi^2 = 9.72$ for ten degrees of
freedom).  We constrain the evolution of $\log L_{\rm X}$/SFR to be independent
of or at most weakly dependent on redshift [$\propto (1+z)^{0.32 \pm 0.88}$].

Based on radio observations of distant star-forming galaxies with
SFR~$\approx$~\hbox{3--300}~\sfr\ in the \cdfn\ and \cdfs, Barger \etal\ (2007)
reported that the \xray\ upper limits for \xray--undetected sources were below
the level expected from the \Lx-SFR correlaiton, thus suggesting that the
correlation may not hold in the high-redshift universe.  However, our stacking
results suggest that the \Lx-SFR correlation {\it does} in fact hold for
average galaxies with SFR~=~\hbox{1--5}~\sfr\ (SFR~=~\hbox{15--100}~\sfr) out
to $z \approx 0.5$ ($z \approx 1.4$).  

For illustrative purposes we have created Figure~11, which shows $\log$~SFR
versus $\log$~\Lx\ for normal star-forming galaxies selected from several
different sources including \xray--detected galaxies from our main sample ({\it
filled gray circles and limits\/}), stacked galaxies from this study ({\it
filled black symbols\/}), local galaxies from PR07 ({\it open circles\/}),
local ultraluminous infrared galaxies from PR07 (ULIRGs; {\it open squares\/}),
and stacked $z \sim 3$ Lyman break galaxies (LBGs; {\it stars}; see $\S$~6 for
details).  For all data used in this plot, we have normalized SFRs
appropriately to be consistent with our adopted Kroupa~(2001) IMF and have
adjusted \xray\ luminosities to correspond to the \hbox{0.5--8~keV} band using
a $\Gamma=2$ power-law SED.  The best-fit \Lx-SFR correlations for local
galaxies from PR07 ({\it solid curve\/}) and $z =$~0--1.4 late-type galaxies
from this study ($\log L_{\rm X}$/SFR~=~39.87; {\it dotted curve\/}) have been
shown for reference.  For comparison, we have shown the SFRs for the MW
and local galaxies M101, M82, and NGC~3256.

\subsection{AGN Contribution to Stacked Signals}

In this section we estimate the contribution to our stacked signals from
contaminating AGNs that have \xray\ luminosities below our \xray\ detection
limit.  This analysis is similar in nature to that in $\S\S$~3.1 and 3.2.2 of
Lehmer \etal\ (2007), which was performed for early-type galaxies.  We
implement the observed cumulative AGN fraction $f_C$: the fraction of galaxies
harboring an AGN with \hbox{2--8~keV} luminosity of $L_{\rm 2-8~keV}$ or
greater.  Hereafter, we compute $f_C$ by taking the number of candidate AGNs in
a particular galaxy sample with a \hbox{2--8~keV} luminosity of $L_{\rm
2-8~keV}$ or greater and dividing it by the number of galaxies in which we
could have detected an AGN with luminosity $L_{\rm 2-8~keV}$.  The latter
number is computed by considering the redshift of each galaxy and its
corresponding sensitivity limit, as obtained from spatially varying sensitivity
maps (see $\S$~4.2 of Alexander \etal\ 2003 and $\S$~4 of Lehmer \etal\ 2005b);
these sensitivity maps were calibrated empirically using sources detected by
{\ttfamily wavdetect} at a false-positive probability threshold of $1 \times
10^{-5}$.  

The quantity $f_C$ is not only a function of $L_{\rm 2-8~keV}$, but is also
dependent on the selection of the galaxy sample considered: in our case
redshift and the physical property $f_{\rm phys}$ (i.e., $L_B$, $M_\star$, and
SFR) of the galaxy population may plausibly play a role in $f_C$.  In our
analyses, we assume that each of the respective dependencies (i.e., $L_{\rm
2-8~keV}$, $z$, and $f_{\rm phys}$) are independent of each other, such that
$f_C \propto g_L(L_{\rm 2-8~keV}) \times g_z(z) \times g_p(f_{\rm phys})$,
where $g_L$, $g_z$, and $g_p$ represents the functional dependence of the
cumulative AGN fraction for each indicated variable $L_{\rm 2-8~keV}$, $z$, and
$f_{\rm phys}$, respectively.  We made use of the \hbox{2--8~keV} bandpass
because of its ability to probe relatively unattenuated \xray\ emission in a
regime of the \xray\ spectrum where we expect there to be minimal emission from
normal galaxies (see also criterion~1 of $\S$~4.1 for further details).  In
total 62 ($\approx$51\%) of our 121 \xray--detected AGN candidates had
\hbox{2--8~keV} detections; we use these AGNs in our AGN fraction analyses.  

%
%
\begin{figure}
\figurenum{13}
\centerline{
\includegraphics[width=9.3cm]{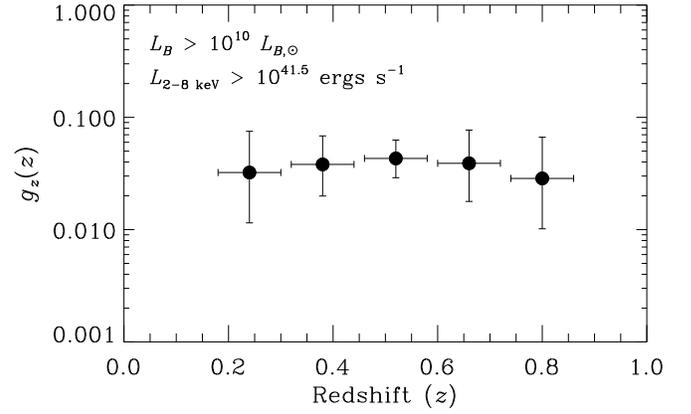}
}
\caption{Cumulative redshift-dependent AGN fraction $g_z(z)$ versus redshift
for late-type galaxy samples with $L_B \simgt 10^{10}$~\lbsol\ and $L_{\rm
2-8~keV} \simgt 10^{41.5}$~\xlum.  We find no significant evolution of $g_z(z)$
over the redshift range \hbox{$z \approx$~0.1--0.8}.
}
\vspace{-0.15in}
\end{figure}

%
%

\begin{figure*}
\figurenum{14}
\centerline{
\includegraphics[width=18.3cm]{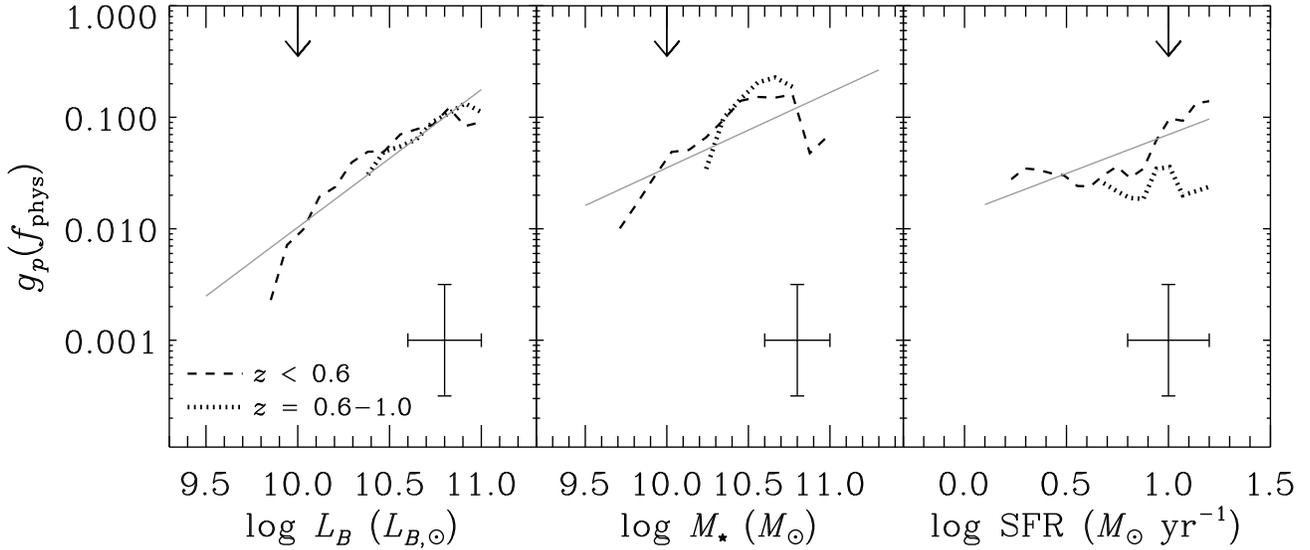}
}
\caption{Cumulative AGN fraction $g_p(f_{\rm phys})$ versus ({\it a\/}) $L_B$,
({\it b\/}) $M_\star$, and ({\it c\/}) SFR for late-type galaxies with $L_{\rm
2-8~keV} \simgt 10^{41.5}$~\xlum\ at $z < 0.6$ ({\it dashed curves\/}) and
\hbox{$z =$~0.6--1} ({\it dotted curves\/}).  In each plot, we have indicated the
best-fit relation ({\it solid line\/}), which is calculated using all galaxies
with $L_{\rm 2-8~keV} \simgt 10^{41.5}$~\xlum\ and $z < 0.6$.  The
downward-pointing arrows indicate which values of $f_{\rm phys}$ were chosen
for computing $g_0$ in equation~8.  For reference, the bin sizes of $f_{\rm
phys}$ used to compute each value of $g_p(f_{\rm phys})$ and the typical
errors of $g_p(f_{\rm phys})$ have been indicated in the lower right-hand
corner of each plot.}
\vspace{-0.15in}
\end{figure*}

We began constructing $f_C$ by estimating the shape of $g_L(L_{\rm 2-8~keV})$
using late-type galaxies with \hbox{$z =$~0--1} and $L_B \simgt 2 \times
10^{10}$~\lbsol, an optical luminosity regime where we have a relatively large
number of sources and are sufficiently complete out to $z = 1$ (see Fig~4$a$).
We split this sample into two subsets about $z = 0.6$, to test whether there is
substantial evolution in the shape and normalization of $g_L(L_{\rm 2-8~keV})$
over this redshift range.  In Figure~12, we show $g_L(L_{\rm 2-8~keV})$ for
galaxies in the redshift ranges $z \simlt 0.6$ ($z_{\rm median} = 0.51$; {\it
filled circles\/}) and \hbox{$z =$~0.6--1} ($z_{\rm median} = 0.84$; {\it open
triangles\/}).  From Figure~12, we see that the overall shape and normalization
of $g_L(L_{\rm 2-8~keV})$ for late-type galaxies with $L_B\simgt 2 \times
10^{10}$~\lbsol\ is similar for galaxies at $z_{\rm median} = 0.51$ and $z_{\rm
median} = 0.84$.  We fit the shape of $g_L(L_{\rm 2-8~keV})$ using least-squares
fitting of the \hbox{2--8~keV} luminosity dependent cumulative AGN fraction
using all galaxies from \hbox{$z =$~0--1} with $L_B\simgt 2 \times
10^{10}$~\lbsol.  For these fits, we found that the data were well-fit by an
exponential function, which we parameterized as $\log g_L(L_{\rm 2-8~keV})
\propto$~\hbox{$a \exp[-b(\log L_{\rm 2-8~keV}-39)^2]$}, where $a$ and $b$ are
fitting constants.  By construction, this function is only valid for $\log
L_{\rm 2-8~keV} > 39$, which is $\approx$1--2 orders of magnitude less luminous
than a typical stacked \xray\ luminosity of our late-type galaxy samples (see
Table~2 and Fig.~10).  We find best-fit values of $a=-0.8$ and $b=-0.05$; in
Figure~12 ({\it thick black curve\/}), we show our best-fit relation for
$g_L(L_{\rm 2-8~keV})$.

We constrained further the redshift evolution of $f_C$ [i.e., the shape of
$g_z(z)$] by dividing our main late-type galaxy sample into five nearly
independent redshift bins (with \hbox{$z =$~0.1--0.8}) and calculating $f_C$
for fixed ranges of $L_{\rm 2-8~keV}$ and $f_{\rm phys}$.  In Figure~13, we
show $g_z(z)$ as a function of redshift for late-type galaxies with $L_{\rm
2-8~keV} \simgt 10^{41.5}$~\xlum\ and $L_B \simgt 10^{10}$~\lbsol, the
approximate completeness limit at $z \approx 0.8$ (see Fig.~4$a$).  Using these
data and $\chi^2$ fitting [assuming a $(1+z)^n$ dependence], we constrained the
redshift evolution of $g_z(z)$ to be proportional to $(1+z)^{0.15 \pm 0.97}$;
similar results were found for different ranges of $L_{\rm 2-8~keV}$ and $L_B$.
This result differs from that observed for early-type galaxies, where the AGN
fraction and mean AGN emission has been found to evolve as $\approx$$(1+z)^3$
(e.g., Brand \etal\ 2005; Lehmer \etal\ 2007).\footnote{We note that the lack
of evolution of the AGN fraction does {\it not} necessarily imply that the
optically-luminous (i.e., $L_B \simgt 10^{10}$~\lbsol) late-type galaxy AGN
number density is constant over \hbox{$z \approx$~0--1}.  For example, $L_B^*$
has been shown to fade by $\approx$1~mag since $z \approx 1$ (e.g., Lilly
\etal\ 1995; Wolf \etal\ 2003; Faber \etal\ 2007), suggesting that there are
fewer optically luminous late-type galaxies in the local universe than at $z =
1$.} We therefore conclude that there is little redshift evolution in the
late-type galaxy AGN fraction over the redshift range \hbox{$z =$~0.1--0.8},
and hereafter we assume that $g_z(z)$ remains roughly constant out to $z =
1.4$.  

To constrain the overall dependence of $f_C$ on $f_{\rm phys}$ [i.e.,
$g_p(f_{\rm phys})$], we calculated the cumulative AGN fractions for late-type
galaxy samples with $f_{\rm phys}$ by holding the ranges of $L_{\rm 2-8~keV}$
and $z$ fixed and varying $f_{\rm phys}$.  In Figure~14, we show $g_p(f_{\rm
phys})$ versus $\log L_B$ (Fig.~14$a$), $\log M_\star$ (Fig.~14$b$), and $\log
{\rm SFR}$ (Fig.~14$c$) for $L_{\rm 2-8~keV} \simgt 10^{41.5}$~\xlum\ at $z
\simlt 0.6$ ({\it dashed curves\/}) and \hbox{$z =$~0.6--1} ({\it dotted
curves\/}).  We calculated $g_p(f_{\rm phys})$ for intervals of $f_{\rm phys}$
where we are approximately complete at $z = 0.6$ (for the $z \simlt 0.6$
interval) and $z = 1$ (for the \hbox{$z =$~0.6--1} interval; see Fig.~4).  For
each sample, we again utilized least-squares fitting to approximate the $f_{\rm
phys}$ dependence of $g_p(f_{\rm phys})$.  These fits were performed using all
data over the redshift range \hbox{$z =$~0--1} assuming a functional dependence
of $\log g_p(f_{\rm phys}) \propto c \log f_{\rm phys}$.  In each panel of
Figure~14, we show the best-fit solutions for $g_p(f_{\rm phys})$ with the gray
lines.  We find that the AGN fraction is strongly dependent on $L_B$
(Fig.~14$a$) and $M_\star$ (Fig.~14$b$), such that more optically-luminous and
massive galaxies have larger AGN fractions; this result is consistent with
other studies of the AGN host galaxies (e.g., Kauffmann \etal\
2003b; Nandra \etal\ 2007; Silverman \etal\ 2007).  Also, the AGN fraction
seems to be mildly dependent on the galaxy SFR; however, this is likely due to
the fact that the SFR is larger on average for more massive galaxies.

Based on the above estimates of the shapes of $g_L(L_{\rm 2-8~keV})$,
$g_z(z)$, and $g_p(f_{\rm phys})$, we approximated empirically $f_C$ for each
choice of $f_{\rm phys}$ following:
\begin{equation}
\log f_C = \log g_0 + a\exp[-b(\log L_{\rm 2-8~keV} - 39)^2] + c\log f_{\rm
phys},
\end{equation}
where $g_0$ represents the normalization of each relation based upon the
value of $f_C$ at $L_{\rm 2-8~keV} > 10^{41.5}$~\xlum\ and \hbox{$L_B = 10^{10}$~\lbsol},
\hbox{$M_\star = 10^{10}$~\msol}, and SFR~$= 2$~\sfr, for galaxy samples selected via
$L_B$, $M_\star$, and SFR, respectively.  We note that only $g_0$ and $c$ are
dependent upon our choice of $f_{\rm phys}$; we find for the set of physical
parameters $f_{\rm phys} =$~[$L_B$, $M_\star$, SFR] that $g_0 =$~[$6.1\times
10^{-14}$, $8.2 \times 10^{-8}$, 0.20] and $c =$~[1.2, 0.67, 0.66].  For
reference, in Figure~12 we have shown curves of $f_C$ for $L_B = 5 \times
10^{9}$, $10^{10}$, and $10^{11}$~\lbsol\ ({\it dotted curves\/}).

To estimate the AGN contamination expected for each of our stacked samples
presented in $\S$~5.1, we followed closely the procedure in $\S$~3.2.2 of
Lehmer \etal\ (2007).  For completeness, we outline this procedure below. 

For each stacked sample, we used equation~8 to compute cumulative AGN fractions
$f_C$.  In Figure~15$a$, we show $f_C$ for two of our $L_B$-selected samples:
$z \approx 0.2$ galaxies with $L_B = 10^{9.5}$--10$^{10}$~\lbsol\ ({\it dashed
curve\/}) and $z \approx 1.3$ galaxies with $L_B =
10^{10.5}$--10$^{11.3}$~\lbsol\ ({\it solid curve\/}).  We then converted $f_C$
for each sample to a differential AGN fraction $f_D$ (i.e., the \hbox{2--8~keV}
luminosity dependent fraction of galaxies harboring AGNs within discrete \xray\
luminosity bins of width $\Delta \log L_{\rm 2-8~keV} = 0.5$; see Fig.~15$b$).
For each sample, we calculated the \hbox{2--8~keV} luminosity dependent
fraction of galaxies that were below our $L_{\rm 2-8~keV}$ detection limit
$f_B$ (Fig.~15$c$).  We then multiplied $f_D$ and $f_B$ to estimate the
\hbox{2--8~keV} luminosity dependent fraction of galaxies harboring an AGN that
was undetected in the \chandra\ exposures $f_U$ (i.e., $f_U = f_D \times f_B$;
see Fig.~15$d$); these AGNs would {\it not} have been removed for our stacking
analyses.  Finally, we approximated the total \hbox{2--8~keV} AGN contamination
$L_{\rm 0.5-2~keV}({\rm contam})$ of each stacked sample using the following
summation:
\begin{equation}
L_{\rm 2-8~keV}({\rm contam}) = \sum_i f_{U,i} \times L_{{\rm 2-8~keV},i},
\end{equation}
where the summation is over all bins of $\Delta \log L_{\rm 2-8~keV} = 0.5$ in
the range $\log L_{\rm 2-8~keV} =$~39--45.  We converted each value of $L_{\rm
2-8~keV}({\rm contam})$ to estimated values of $L_{\rm 0.5-8~keV}({\rm
contam})$ and $L_{\rm 0.5-2~keV}({\rm contam})$ by assuming the contaminating
AGN emission roughly follows an \xray\ SED described by a power-law.  In order
to constrain the average photon index of the power-law, we stacked all
\xray--detected AGNs in our stacked samples with \hbox{0.5--8~keV} luminosities
below $10^{42}$~\xlum.  For these AGNs, we find a stacked effective photon
index of $\Gamma_{\rm eff} = 0.97 \pm 0.03$.  If the intrinsic value of the
photon index is $\Gamma=2$, at the median redshift of our main sample ($z_{\rm
median} = 0.5$), $\Gamma_{\rm eff} \approx 1$ corresponds to an intrinsic
\xray\ column density of $N_{\rm H} \approx$~\hbox{1--2}~$\times
10^{22}$~cm$^{-2}$.  If we assume $\Gamma_{\rm eff} = 1$ describes well the
effective SED of the \xray--undetected AGNs in our stacked sample, we find that
AGN contamination can account for \hbox{$\approx$1--30\%} (median of
$\approx$5\%) of the \hbox{0.5--8~keV} emission from our stacked samples,
suggesting that AGNs are not providing a significant contribution to our
stacked results.  We note that the \xray\ SED used in this calculation has an
important effect on the overall estimate of the AGN contamination.  Since our
estimates for contamination in the \hbox{0.5--8~keV} and \hbox{0.5--2~keV}
bandpasses decrease as $\Gamma_{\rm eff}$ decreases, the amount of
contamination in our samples may be underestimated if our choice of
$\Gamma_{\rm eff}$ is too flat; however, we find that for conservative choices
of $\Gamma_{\rm eff}$ (i.e., $\Gamma_{\rm eff} \simlt 2$) that are
representative of even unobscured AGNs (for reference, see Fig.~5), the AGN
contamination remains low ($\simlt$40\%) and has no material effect on our
results.

It is also worth noting that if there exists a large population of
intrinsically luminous yet heavily obscured AGNs in our high-redshift galaxy
samples that are \xray--undetected in the CDFs, then our conclusions above could
be somewhat different.  If such a population were present and had significant
influence on our stacked results, then we would expect to find stacked \xray\
spectra that were relatively flat ($\Gamma_{\rm eff} \simlt 1$); however, as
discussed in $\S$~5.1, we find that very few of our stacked samples could have
such flat spectra in our stacked samples, suggesting that such a population, if
present, does not have a strong effect on our results.

%
%
\begin{figure}
\figurenum{15}
\centerline{
\includegraphics[width=9.cm]{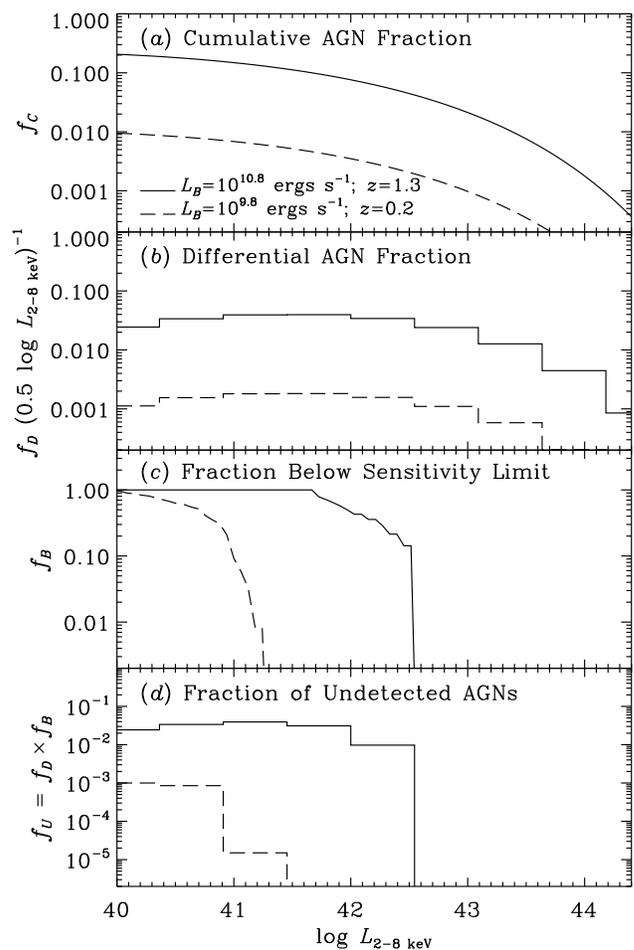}
}
\caption{
\small ({\it a\/}) Cumulative AGN fraction (i.e., the fraction of galaxies
harboring an AGN with a \hbox{2--8~keV} luminosity of $L_{\rm 2-8~keV}$ or
greater), $f_C$, versus $\log L_{\rm 2-8~keV}$ for two of our stacked samples:
$z \approx 0.2$ galaxies with $L_B = 10^{9.5}$--10$^{10}$~\lbsol\ ({\it dashed
curve\/}) and $z \approx 1.3$ galaxies with $L_B = 10^{10.5}$--10$^{11.3}$~\lbsol\
({\it solid curve\/}).  Each curve was computed following equation~8.  ({\it
b\/}) Differential AGN fractions (i.e., the fraction of galaxies harboring an AGN
in discrete bins of width $\Delta \log L_{\rm 2-8~keV}=0.5$), $f_D$, versus
$\log L_{\rm 2-8~keV}$.  ({\it c\/}) Fraction of late-type galaxies for which we
could {\it not} have detected an AGN with a \hbox{2--8~keV} luminosity of
$L_{\rm 2-8~keV}$, $f_B$, versus $\log L_{\rm 2-8~keV}$.  ({\it d\/}) Fraction of
galaxies harboring AGNs in our optically luminous faded samples that would
remain undetected due to sensitivity limitations, $f_U = f_D \times f_B$,
versus $\log L_{\rm 2-8~keV}$; these galaxies would not have been removed from
our stacking analyses. 
\vspace{-0.15in}
}
\end{figure}

In Table~2 (cols.~25--27), we have provided the estimated fractional AGN
contribution to each stacked sample [i.e., $L_{E_1-E_2}({\rm
contam})/L_{E_1-E_2}$] for the FB, SB, and HB using the technique described
above and an assumed $\Gamma_{\rm eff}=1$.  We find that the estimated AGN
contamination is most significant for galaxy samples with large values of $L_B$,
$M_\star$, and SFR.

\section{Extension to Distant Lyman Break Galaxies}

As noted in $\S$~1, the global star-formation rate density has been observed to
increase with redshift out to \hbox{$z \sim$~1--1.5}.  Investigations of the
most-distant LBGs at \hbox{$z \sim$~2--7} show that the
star-formation density peaks around \hbox{$z \approx$~1--3} and gradually
declines toward higher redshifts (e.g., Steidel \etal\ 1999; Giavalisco 2002,
2004b; Bouwens \etal\ 2004, 2005; Dickinson \etal\ 2004).  To investigate
whether the mean \xray\ activity from normal late-type galaxies follows a
similar trend, we study the \xray\ properties of a sample of $z = 3.01 \pm
0.24$ LBGs, which were identified as $U$-band ``dropouts'' through the GOODS
project (see Lehmer \etal\ 2005a for details).  

We filtered the GOODS $z \sim 3$ LBG sample (Lee \etal\ 2006) to include only
LBGs that (1) were within the central $\approx$4\farcm0 of the $\approx$2~Ms
\cdfn\ and $\approx$1~Ms \cdfs, and (2) had rest-frame $B$-band luminosities
that were similar to the most optically luminous ($L_B =$~3--20~$\times
10^{10}$~\lbsol) \hbox{$z =$~0--1.4} late-type galaxies used in this study.
$B$-band luminosities were calculated by applying $k$-corrections to the
$z_{850}$-band flux (from GOODS), where $k$-corrections were derived using an
SED appropriate for LBGs (see $\S$~2.2 of Lehmer \etal\ 2005a for details).  We
found that 85 $z \sim 3$ LBGs from the Lehmer \etal\ (2005a) sample satisfied
these two selection criteria.  

We identified three \xray--detected LBGs, which all had $L_{\rm 2-8~keV} \simgt
10^{43}$~\xlum.  Due to their high \xray\ luminosities, we classified these
sources as obvious AGNs.  After removing these three \xray--detected AGNs from
our $z \sim 3$ LBG sample, we performed \xray\ stacking analyses as described
in $\S$~4.2.  We found a significant ($3.7\sigma$) detection in the
\hbox{0.5--2~keV} bandpass, which corresponds roughly to rest-frame
\hbox{2--8~keV} emission.  Assuming an intrinsic power-law \xray\ spectrum with
a photon index of $\Gamma = 2$, we found a mean \hbox{0.5--8~keV} luminosity of
\hbox{$L_{\rm X} = (4.9 \pm 1.4) \times 10^{41}$~\xlum} for our $z \sim 3$
LBGs, a value that agrees well with previous studies (e.g., Lehmer \etal\
2005$a$; Laird \etal\ 2006).

%
%
\begin{figure}
\figurenum{16}
\centerline{
\includegraphics[width=9.cm]{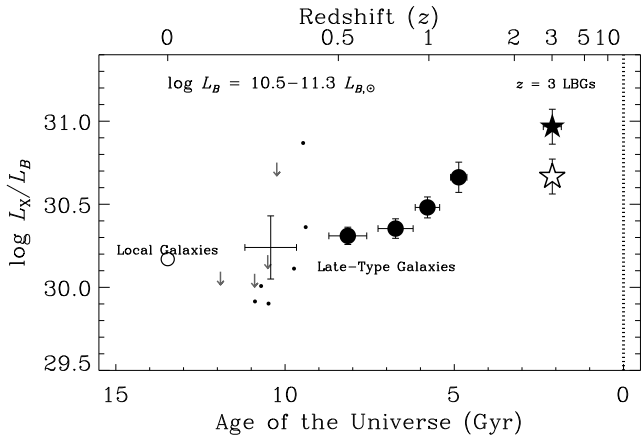}
}
\caption{
Logarithm of the \xray--to--$B$-band mean luminosity ratio $\log L_{\rm X}/L_B$
versus the age of the Universe (for our adopted cosmology, the current age of
the Universe is 13.47~Gyr) for star-forming galaxies with $L_B =$~\hbox{3--20}~$\times
10^{10}$~\lbsol.  For reference, redshift has been plotted along the top axis.
We have included mean values of $L_{\rm X}/L_B$ for the S01 local sample of
late-type galaxies ({\it open circles\/}), our stacked samples at $z
\approx$~\hbox{0.5--1.4} ({\it large filled circles\/}), and $z \sim 3$ LBGs that were
both uncorrected ({\it filled star\/}) and corrected ({\it open star\/}) for
AGN contamination.  At $z \approx$~\hbox{0.1--0.4}, we have plotted $L_{\rm X}/L_B$
for individual late-type galaxies from our main sample ({\it small filled
circles} and {\it upper limits\/}); the mean values and errors for these
galaxies, computed using {\ttfamily ASURV}, has been indicated.  
\vspace{-0.15in}
}
\end{figure}

Since our LBGs reside in the high-redshift universe, AGN contamination is
expected to have a more significant effect on these results than it did for our
\hbox{$z =$~0--1.4} late-type galaxies.  The median \xray\ luminosity detection
limit is \hbox{$\approx$$8 \times 10^{41}$~\xlum} and $\approx$$10^{43}$~\xlum\
for the \hbox{0.5--2~keV} and \hbox{2--8~keV} bandpasses, respectively.  To
estimate the AGN contamination, we followed the approach outlined in $\S$~5.2,
which made use of the \hbox{2--8~keV} luminosity dependent AGN fraction
($f_C$).  In Figure~12 ({\it filled stars\/}), we show the \xray\ luminosity
dependent AGN fraction $g_{L, \rm LBG}(L_{\rm 2-8~keV})$ for $z\sim3$ LBGs.  We
note that at $L_{\rm 2-8~keV} \simgt 10^{43}$~\xlum, the cumulative AGN
fraction for $z\sim3$ LBGs is a factor of $\approx$3--4 times larger than that
computed for our \hbox{$z =$~0--1} late-type galaxy sample with similar optical
luminosities.  Using the functional form for $f_C$ presented in equation~8, but
with a \hbox{$\approx$3--4} times larger normalization factor for $z \sim 3$
LBGs (i.e., $g_{0, {\rm LBGs}} \approx$~3--4~$g_0$), we find that AGN emission
may plausibly account for \hbox{$\approx$50--70\%} of the stacked
\hbox{0.5--2~keV} counts. 

In Figure~16, we show the \xray--to--$B$-band mean luminosity ratio for $L_B
=$~3--20~$\times 10^{10}$~\lbsol\ star-forming galaxies (i.e., late-type
galaxies and LBGs) as a function of the age of the Universe.  Together, these
data span $\approx$85\% of cosmic history (i.e., out to $z \sim3$).  As
presented in $\S$~5.1, \Lx/$L_B$ shows significant evolution over the redshift
range $z =$~0--1.4, and after correcting for AGN contamination, we find that
\Lx/$L_B$ is similar for $z \sim 3$ LBGs and $z = 1.4$ late-type galaxies.
This result suggests that the non-AGN \xray\ emission for the most luminous
star-forming galaxies may flatten near \hbox{$z \approx$~1.4--3}, which has
been predicted roughly from simulations of how the normal-galaxy \xray\
emission is expected to respond due to global changes in the star-formation
rate density (e.g., Ghosh \& White~2001).

To test whether the \Lx-SFR correlation is similar for \hbox{$z\sim3$} LBGs as
we found for late-type galaxies at \hbox{$z =$~0--1.4}, we approximated
absorption-corrected SFRs for the $z\sim3$ LBGs using UV band emission.  These
SFRs were approximated following SFR~=~$9.8 \times 10^{-11} \gamma L_{\rm UV}$,
where \hbox{$\gamma \approx 6$} is the absorption-correction factor (see
Giavalisco \etal\ 2004b).  As described in $\S$~3.3, we approximated the UV
luminosity following $L_{\rm UV} = 3.3 \nu l_\nu$(2800~\AA); however, here
$l_\nu$(2800~\AA) was derived using our adopted LBG SED.  We find that the mean
SFR for optically-luminous $z\sim3$ LBGs is \hbox{$\approx$60~\sfr}.  After
correcting the mean stacked \xray\ luminosity for AGN contamination, we find an
\xray--to--SFR ratio of $\log L_{\rm X}$/SFR~$\approx$~\hbox{39.4--39.6}, which
is suggestively lower than its value for \hbox{$z =$~0--1.4} late-type galaxies
(i.e., $\log L_{\rm X}$/SFR~\hbox{$= 39.87 \pm 0.04$}).  We note, however, that
the estimation of the mean SFR for these galaxies is based solely on the
2800~\AA\ emission and is therefore highly uncertain; further constraints on
the mean infrared luminosity of these sources would help considerably.  For
reference, in Figure~11 we have plotted $\log$~SFR versus $\log L_{\rm X}$ for
the \hbox{$z\sim3$} LBGs both uncorrected ({\it filled star\/}) and corrected
({\it open star\/}) for AGN contamination.  

It is interesting to note that the AGN-corrected LBG sample appears to have
$L_{\rm X}$ and SFR values similar to those of local ULIRGs, which are
relatively \xray\ underluminous for their derived SFRs.  PR07 suggest that
galaxies with SFRs that begin to exceed $\simgt$50--100~\sfr, similar to the
ULIRGs, may have \xray\ properties that are completely dominated by HMXBs.  In
such systems, other \xray--emitting populations that may normally be
significant in more quiescent galaxies (e.g., LMXBs, hot gas, supernovae and
their remnants, etc.) would collectively contribute only a negligible fraction
of the total \xray\ emission, thus causing the overall galaxies to appear
\xray\ underluminous compared with the \Lx--SFR correlation.  We note, however,
that our results presented in $\S$~5.1 suggest that HMXBs likely dominate the
X-ray emission from galaxies with SFRs much lower than those of ULIRGs.
Another possibility is that as the absorption within the most actively
star-forming galaxies increases with SFR (see discussion in $\S$~5.1), the
\xray\ emission from point-source populations may become significantly
obscured.  Deep \xray\ observations of local LIRGs and ULIRGs with \chandra\
and \xmm\ have found significant \xray\ absorption in their point-source
populations (see, e.g., Lira \etal\ 2002; Zezas \etal\ 2002, 2003) thus
providing some support for this possibility.  Future investigations into the
nature of the \xray\ populations of ULIRGs could help resolve this issue.

\section{Summary and Future Work}

We have investigated the \xray\ emission from 2568 normal late-type galaxies
over the redshift range $z =$~0--1.4 that lie within the \chandra\ Deep Fields
(CDFs).  Our late-type galaxy sample was constructed primarily using
color-magnitude diagrams, which incorporated rest-frame $U-V$ color and
absolute $V$-band magnitudes, to isolate blue late-type galaxies (see $\S$~2
for details).  In total, 225 ($\approx$9\%) of our late-type galaxies were
detected individually in the \xray\ band.  Based on \xray\ and optical spectral
properties, \xray--to--optical flux ratios, the correlation between \xray\
luminosity and star-formation rate, and comparisons between infrared and radio
properties, we infer that 121 ($\approx$53\%) of the \xray--detected late-type
galaxies are dominated by AGN emission.  The remaining 104 \xray--detected
galaxies had \xray\ and multiwavelength properties consistent with those of
normal late-type galaxies with \xray\ emission dominated by \xray\ binaries
(HMXBs and LMXBs).  To study the \xray\ emission and evolution from large
representative populations of late-type galaxies (i.e., including the
\xray--undetected sources), we utilized \xray\ stacking analyses of galaxy
populations with AGN candidates removed.  We stacked normal-galaxy samples that
were selected via their rest-frame $B$-band luminosity ($L_B$), stellar mass
($M_\star$), and star-formation rate (SFR) in redshift bins (see $\S$~5.1).
Furthermore, we compared these results with those found for a sample of
$z\sim3$ LBGs from Lehmer \etal\ (2005a).  In the points below, we summarize
our key findings:

\vspace{0.15in}

1. We obtained significant detections in the \hbox{0.5--2~keV} and
\hbox{0.5--8~keV} bandpasses for all of our stacked samples.  We estimated that
LMXBs and low-level AGNs provide only low-level contributions to the stacked
\xray\ emission from our samples and that HMXBs constitute the dominant
\xray--emitting component.  Therefore, for these galaxies the \xray\ emission
is tracing primarily star-formation activity.  Normal late-type galaxy samples
selected via $L_B$ and $M_\star$ show significant (at the $>$99.9\% confidence
level) evolution in their average \xray\ properties from $z = 0$ to 1.4.  For
the most optically luminous ($L_B \approx 5\times 10^{10}$~\lbsol) and massive
(\hbox{$M_\star \approx 3 \times 10^{10}$~\msol}) late-type galaxies at
$z=1.4$, $L_{\rm X}/L_B$ and $L_{\rm X}/M_\star$ are measured to be larger than
their local values by factors of \hbox{$\approx$2--4} and
\hbox{$\approx$7--13}, respectively.

2. We find that late-type galaxies of lower stellar mass generally have larger
\xray--to--stellar-mass mean ratios ($L_{\rm X}/M_\star$) than their
higher-mass analogs.  Over $z \approx$~0.2--1, galaxies with $M_\star \approx 6
\times 10^{9}$~\msol\ are a factor of \hbox{$\approx$2--3} times more \xray\
luminous per unit stellar mass than galaxies with $M_\star \approx 3 \times
10^{10}$~\msol.  

3. We characterized the \xray\ properties of 888 24$\mu$m-detected late-type
galaxies in our sample.  The 24$\mu$m data allowed us to select galaxy samples
selected via their SFRs.  For these samples, we found that the \xray\
luminosity is well-predicted by a constant \Lx--to--SFR ratio, similar to the
\Lx-SFR correlation reported by previous authors (e.g., PR07).  This implies
that the \Lx-SFR correlation holds out to at least $z = 0.5$, 1, and 1.4 for
galaxies with SFR~$\approx$~2, 10, and 50~\sfr, respectively, and supports the
idea that the strong \xray\ evolution observed for normal late-type galaxies
selected via $L_B$ and $M_\star$ is likely due to strong changes in SFR.

4. The \xray\ properties of our most optically-luminous ($L_B =$~3--20~$\times
10^{10}$~\lbsol) late-type galaxies at $z=1.4$ are comparable to those for $z
\sim 3$ LBGs with similar optical luminosities, once \xray--undetected AGN
contamination in the LBG population has been accounted for.  This suggests that
there may plausibly be a flattening in the $L_{\rm X}/L_B$--$z$ relation for
optically-luminous star-forming galaxies between $z \sim$~1.4--3.  We estimate
a mean SFR of $\approx$60~\sfr\ for these LBGs.  We find that the observed mean
\xray\ luminosity is suggestively underluminous based on the \Lx-SFR
correlation prediction; this result is similar to that found for local ULIRGs
with comparable SFRs. \\

The above results can be improved greatly through (1) the study of late-type
galaxy populations in other existing multiwavelength extragalactic survey
fields that contain \chandra\ observations, (2) additional observations of the
CDFs to even better sensitivity levels than are currently available, and (3)
observations with future \xray\ missions with imaging capabilities that are
complementary to those of \chandra.

The areal footprint of the CDF regions used in this study totals
$\approx$0.18~deg$^2$ (see $\S$~2.1 for details), which severely limits the
number of galaxies being studied at $z \simlt 0.2$.  The redshift range
\hbox{$z =$~0--0.2} spans $\approx$2.5~Gyr of cosmic look-back time, compared
with the $\approx$6.6~Gyr over \hbox{$z =$~0.2--1.4}, that we are most
effectively studying above.  Several complementary deep and wide extragalactic
\chandra\ survey fields that have recently been conducted or are in progress
can improve the present situation.  Three such ideal survey fields are the
$\approx$200~ks All-wavelength Extended Groth Strip International Survey
(AEGIS; e.g., Nandra \etal\ 2005; Davis \etal\ 2006), the $\approx$50~ks
\chandra\ Cosmic Evolution Survey (C-COSMOS; PI: M.~Elvis; see also, Scoville
\etal\ 2007),\footnote{We note that the C-COSMOS observing strategy has
been designed to have several overlapping $\approx$50~ks ACIS-I exposures; in
these regions, the total exposure will reach $\approx$200~ks due to the
overlaping exposures.  However, the analyses presented in this paper utilize
only high-quality \chandra\ imaging at off-axis angles of $\simlt$6\farcm0 (see
$\S$~2.1 for additional justification), where there is very little overlap in
the C-COSMOS \chandra\ pointings.  We therefore do not consider the overlapping
regions of the C-COSMOS exposures in this discussion.  For additional
information regarding C-COSMOS, see
http://cfa-www.harvard.edu/hea/cos/C-COSMOS.html.} and the $\approx$5~ks NOAO
Deep Wide-Field Survey (NDWFS; e.g., Murray \etal\ 2005).  

Our study shows that identifying and removing contaminating AGNs is a crucial
ingredient to studying the stacked \xray\ properties of normal-galaxy samples
(see $\S$~5.2).  In order to avoid significant levels of AGN contamination in
shallower wide-area \chandra\ surveys such as AEGIS, C-COSMOS, and NDWFS,
\xray\ studies of normal galaxy populations must be limited to lower redshift
intervals.  If we require a detection limit of $L_{\rm 0.5-2~keV} =
10^{41.5}$~\xlum\ over regions of each survey where the \chandra\ sensitivity
is optimal (i.e., $\simlt$6\farcm0 from the \chandra\ aim-points), then we
estimate that AEGIS, C-COSMOS, and NDWFS could effectively be used to study
normal late-type galaxy populations at redshifts less than $\approx$0.5,
$\approx$0.25, and $\approx$0.1, respectively.  When factoring in the areal
coverage of these fields, we estimate that AEGIS and C-COSMOS will contain
$\approx$250--350 late-type galaxies with \hbox{$L_B \simgt 3 \times
10^9$~\lbsol} at \hbox{$z =$~0.1--0.2} (i.e., a factor of $\approx$3 larger
than the CDFs) and that AEGIS, C-COSMOS, and NDWFS taken together will have
$\approx$100--250 similar galaxies at $z \simlt 0.1$ (i.e., a factor of
$\approx$22 times larger than the CDFs).  Studying these galaxy populations
would significantly improve constraints on the \xray\ properties of $z \simlt
0.2$ late-type galaxies.

Deeper \chandra\ observations of the CDFs would provide additional insight into
the \xray\ properties of the normal late-type galaxy populations presented in
this study and enable us to extend our analyses to higher redshifts.  Typical
galaxies in our sample have mean \xray\ luminosities of \hbox{$L_{\rm X}
\approx 10^{39.5}$--$10^{40}$~\xlum}.  In the $\approx$2~Ms \cdfn, where our
\chandra\ sensitivity is greatest, we expect that galaxies with $L_{\rm X}
\approx 10^{39.5}$~\xlum\ should be detectable out to $z \approx 0.2$.  Out of
the 11 late-type galaxies with $L_B \simgt10^9$~\lbsol\ at $z \simlt 0.2$, we
find \xray\ detections for 8 of them ($\approx$72\%).  Deeper \chandra\
observations over the CDFs will allow for the detection of such galaxies out
to progressively higher redshifts.  For \chandra\ exposures of $\approx$5~Ms
and $\approx$10~Ms, a source with $L_{\rm X} \approx 10^{39.5}$~\xlum\ should
be detectable out to $z \approx 0.3$ and $z \approx 0.4$, respectively.  If we
pessimistically assume an \xray\ detection fraction of $\approx$70\%, we
estimate that $\approx$30 and $\approx$50 normal galaxies would be detected
individually for each respective deep exposure.  The individual detection of
these galaxies would allow for improved constraints on both the \xray\
luminosity functions of normal late-type galaxies at higher redshifts and the
low-luminosity AGN contributions to our stacked signals.  At higher redshifts,
an $\approx$5~Ms and $\approx$10~Ms \chandra\ exposure would enable us to
effectively perform studies of normal late-type galaxies, similar to those
presented in this paper, out to $z \approx 2$ and $z \approx 2.5$,
respectively.  Such a data set would provide, for the first time, a reliable
constraint on the \xray\ emission from normal star-forming galaxies near the
peak of the global star-formation rate density at $z \approx$~1.5--3.

In addition to the improvement that additional \chandra\ observations could
provide, future \xray\ missions such as \xeus\ and \genx\footnote{For
additional information regarding the future \xray\ missions \xeus\ and \genx,
see http://www.rssd.esa.int/XEUS/ and http://genx.cfa.harvard.edu,
respectively.} should enable new scientific investigations of distant normal
galaxies.  \xeus\ will be able to place significant spectral constraints for
sources with \hbox{0.5--2~keV} fluxes down to $\approx$$10^{-17}$~\flux, a
level fainter than many of the average fluxes derived from our samples.  \genx\
is planned to provide imaging with $\approx$0\farcs1 per resolution element and
will easily probe to flux levels of $\approx$$10^{-19}$~\flux\ ($L_{\rm X}
\approx 5 \times 10^{38}$~\xlum\ at $z = 1$).  At these levels, the first
detailed investigations of the evolution of the normal late-type galaxy \xray\
luminosity function can be performed effectively out to $z \simgt 1$ without
confusion problems due to the crowding of large numbers of sources.
Furthermore, these observations will allow for new constraints to be placed on
the \xray\ populations of all normal galaxies in the observable universe that
are offset by more than $\approx$0.8~kpc from their host-galaxy nuclei.

\acknowledgements

We thank Robin Ciardullo, Doug Cowen, Mike Eracleous, Caryl Gronwall, Bin Luo,
Brendan Miller, Tim Roberts, and Ohad Shemmer for useful suggestions, which
have improved the quality of this paper.  We also thank the referee for
carefully reviewing the manuscript and providing detailed comments.  We
gratefully acknowledge financial support from \chandra\ X-ray Center grant
G04-5157A (B.D.L., W.N.B., A.T.S.), the Science and Technology Facilities
Council fellowship program (B.D.L), the Royal Society (D.M.A.), the Emmy
Noether Program of the Deutsche Forschungsgemeinscaft (E.F.B.), NASA LTSA Grant
NAG5-13102 (D.H.M.), the \chandra\ Fellowship program (F.E.B.), and NSF grant
AST 06-07634 (D.P.S.).

%
{}
%

\end{document}